# Imaging over an unlimited bandwidth with a single diffractive surface

*Sourangsu Banerji,* [1,*] *Monjurul Meem,* [1,*] *Apratim Majumder,* [1] *Berardi Sensale-Rodriguez*[1] *and Rajesh Menon*[1, 2, a)]

[1]Department of Electrical and Computer Engineering, University of Utah, Salt Lake City, UT 84112, USA.

[2]Oblate Optics, Inc. San Diego CA 92130, USA.

a) rmenon@eng.utah.edu

[*] Equal contribution.

**ABSTRACT**

It is generally thought that correcting chromatic aberrations in imaging requires multiple surfaces. Here, we show that by allowing the phase in the image plane of a flat lens to be a free parameter, it is possible to correct chromatic aberrations over an almost unlimited bandwidth with a single diffractive surface. Specifically, we designed, fabricated and characterized a flat multi-level diffractive lens (MDL) that images at the wavelengths from 450nm to 850nm. We experimentally characterized the focusing efficiency, modulation-transfer function, wavefront aberrations, vignetting, distortion and signal-to-noise ratio performance of a camera comprised of this MDL and a conventional image sensor. Further, we designed two MDLs with operating wavelengths from 500nm to 15μm, and from 2μm to 150μm, respectively. With no apparent limitation in the operating bandwidth, such flat lenses could replace multiple refractive surfaces that are traditionally required for chromatic corrections, leading to thinner, lighter and simpler imaging systems with bandwidth limited primarily by the quantum efficiency of the sensor.

## Introduction

The lens is considered the most fundamental element for imaging. Imaging is information transfer from the object to the image planes. This can be accomplished via a conventional lens that essentially performs a one-to-one mapping [1], via an unconventional lens (such as one with a structured point-spread function or PSF) that performs a one-to-many mapping, or via no lens, where the light propagation essentially performs a one-to-all mapping. In the first case, the image is formed directly. In the second case, the image is formed after a computation and can be useful, when encoding spectral [2-3] or depth [4] or polarization [5] or other information into the geometry of the PSF itself. Note that the modification of the PSF may be at the same scale as the diffraction limit [2-5] or it can even be much larger [6-9]. The image can be recovered in many cases in the optics-less scenario as well [10,11], but more importantly, machine learning may be employed to make inferences based on the acquired information (even without performing image reconstruction for human visualization), which has potential implications for privacy among other interesting outcomes [12]. However, the conventional lens approach is preferred in many cases due to the high signal-to-noise ratio achievable at each image pixel (resulting from the 1:1 mapping). When this conventional lens is illuminated by a plane wave, it forms a focused spot at a distance equal to its focal length.

Now, if we appeal to the fact that in the vast majority of imaging applications, only the intensity is measured, the phase of the field in the image or focal plane is a free parameter, something which comes from the inverse diffraction transform [13]. Then, it is easy to see that the phase of the plane wave after it transmits the lens can have multiple forms. We extended this concept to a continuous spectrum of wavelengths from 450nm to 850nm, and designed a single Multi-level Diffractive Lens (MDL) with focal length of 1mm and aperture of 150μm, whose schematic is shown in Fig. 1A. The MDL has a fixed geometry as illustrated in Fig. 1B. However, its material dispersion and wavelength determine the phase transmittance function as shown in Figs. 1C-F for λ=500nm, 600nm, 700nm and 800nm, respectively. When these phase distributions are propagated to the focal plane (1mm away), we obtain the phase, amplitude and intensity distributions at λ =500nm (Figs. 1G-I), 600nm (Figs. 1J-L), 700nm (Figs. 1M-O) and 800nm (Figs. 1P-R), respectively. Note that the intensity distributions for all the four wavelengths are almost identical, but their phase distributions are quite different. This example, as well as our argument above simply shows that the lens does not possess a *unique* phase transmittance function. This is in contradiction to most textbooks that teach us that the ideal lens should have a parabolic phase profile. By removing this restriction, we enable numerous solutions to the ideal lens. Then, the final choice can be made based upon other requirements such as achromaticity, minimization of aberrations, manufacturability, etc. Stating this differently, we can show that a single surface can correct chromatic and other imaging aberrations, something that was considered impossible in refractive lenses [14]. In fact, we show that the achievable bandwidth is in fact unlimited for practical purposes. In this paper, we experimentally demonstrate such an example of a single surface, whose phase-transmittance function is engineered to minimize chromatic aberrations over a bandwidth of 450nm to 850nm. We further show via simulations two separate MDLs, each representing a single surface that is engineered to focus light from $\lambda = 0.5\mu m$ to $15\mu m$, and from $\lambda = 2.5\mu m$ to $150\mu m$, respectively.

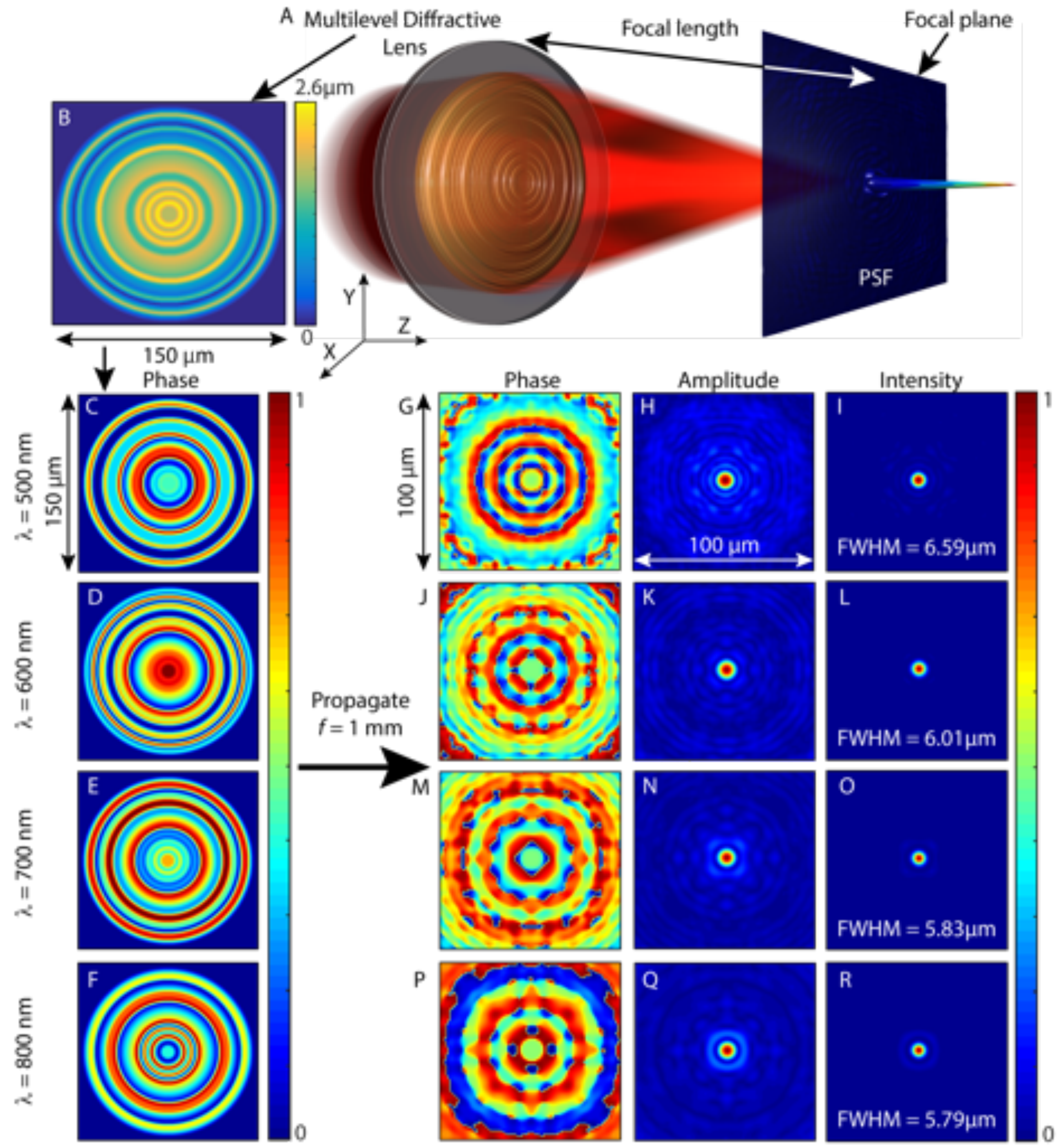

***Figure 1**: (A) Schematic of a single surface multi-level diffractive lens (MDL) that performs close to 1:1 mapping over a broad spectrum. (B) Topography of an MDL with focal length=1mm, aperture=150µm and operating wavelengths from 450nm to 850nm. The minimum feature width, maximum feature height and number of levels for this MDL are 3µm, 2.6µm and 100, respectively. (C) Phase distribution in the MDL plane at λ =500nm. (D-F) The amplitude, phase and intensity distributions in the focal plane for λ =500nm. The corresponding plots for λ =600nm, 700nm and 800nm are shown in (G-R). The phase is normalized to 2π and amplitude and intensity are normalized to their maximum values in all plots. Note that the intensity distributions are almost identical for all four wavelengths, but their phase distributions are quite different. Achromaticity with a single surface is achieved by allowing phase in the focal plane to be a free parameter.*

It is important to note that in conventional refractive imaging systems, multiple lenses (sometimes made from different materials with differing dispersion properties) are used to correct for chromatic aberrations [14]. Not only are the individual refractive lenses thick and heavy, but multiple lenses require precise alignment during assembly. Metalenses have recently become popular to mitigate these disadvantages. By engineering deep subwavelength features on a surface, the parabolic phase profile with group-velocity compensation has been applied to correct for chromatic aberrations [15]. Via a careful literature study (see Table S1) [16], we conclude that the best achromatic metalens in the visible band has demonstrated efficiency of about 35% from 460nm to 700nm for unpolarized light [15]. However, this metalens requires 50nm-wide, 600nm-tall structures in $TiO_2$ because high refractive index is necessary for its operation. Needless to say, no metalens to date has ever reported polarization-insensitive imaging with high efficiency over the visible and near-IR bands. In fact, we recently showed that appropriately designed multi-level diffractive lenses (MDLs) are not only far easier to fabricate, but they outperform most metalenses, and thereby concluded that metalenses do not offer any advantage for imaging [17]. We have already demonstrated MDLs in the visible (450nm to 750nm) [18, 19], coupling two MDLs for magnification, [20], in the long-wave infrared (8μm to 12μm) [21] and in the terahertz band (1 mm to 3 mm) [22, 23].

In this paper, we exploit the concept of non-unique lens phase functions to experimentally demonstrate MDLs with high efficiency over the visible and near-infrared bands (450nm to 850nm) utilizing relatively large microstructures (3μm wide, 2.6μm tall) fabricated in a polymer. Furthermore, we show via simulations two different MDLs with an even more extended operation bandwidth, one that is achromatic from 0.5μm to 15μm and second from 2μm to 150μm.

**Design**

The details of our design methodology have been reported before [18-23]. To summarize, we maximize the wavelength-averaged focusing efficiency of the MDL by selecting the distribution of heights of the rings that form the MDL (see Fig. 1A). This selection is based upon a modified direct binary search technique. We designed, fabricated and characterized an MDL with focal length and NA of 1mm and 0.075, respectively operating at the wavelength range of 450nm to 850nm. The design had a constraint of at most 100 height levels with a maximum individual height level of 2.6μm and minimum feature width of 3μm. The material dispersion of a positive-tone photoresist, S1813 (Microchem) (Fig. S1) was assumed [16]. The height distribution of the designed MDL is shown in Fig. 1B, while the phase transmittance function at four wavelengths are shown in Figs. 1C-F. As mentioned earlier, when these fields are propagated to the focal plane, 1mm away from the MDL, the resulting amplitude and phase distributions as well as the intensity distributions at the corresponding wavelengths are shown in Fig. 1. Note that even though the phase distribution in the focal plane differs between the wavelengths, the intensity distributions are almost identical to that expected from the diffraction-limited case. As a result, we have a single-surface lens that is achromatic from 450nm to 850nm. The simulated PSFs for all the wavelengths are depicted in Fig. S4 [16].

**Experiments**

The MDL was fabricated using grayscale lithography as has been reported previously (Fig. S6) [16,18-21]. An optical micrograph of the fabricated device is shown in Fig. 2A. The point-spread function (PSF) of the MDL was measured by illuminating it by a collimated beam from a tunable supercontinuum source (NKT Photonics SuperK Extreme with SuperK VARIA filter for visible wavelengths, 350nm-850nm and SuperK SELECT filter for near infrared wavelengths, 800nm-1400nm) (Fig. S7) [16, 18-20]. The wavelength of the source was tuned from 450nm to 850nm in steps of 50 nm and bandwidth of 10 nm. The focused spot at each wavelength was relayed with magnification (22.22X) onto a monochrome image sensor (DMM 27UP031-ML, Imaging Source). The captured raw images for four wavelengths are shown in Figs. 2B-E, where we note all the wavelengths are close to focus, exhibiting achromatic behavior. The full-width at half-maximum (FWHM) of the measured PSFs are noted in the corresponding images. The diffraction-limited, measured and simulated FWHM as function of wavelength are plotted in Fig. 2F, which shows good agreement. The measured PSFs for all the wavelengths are shown in Fig. S5 [16].

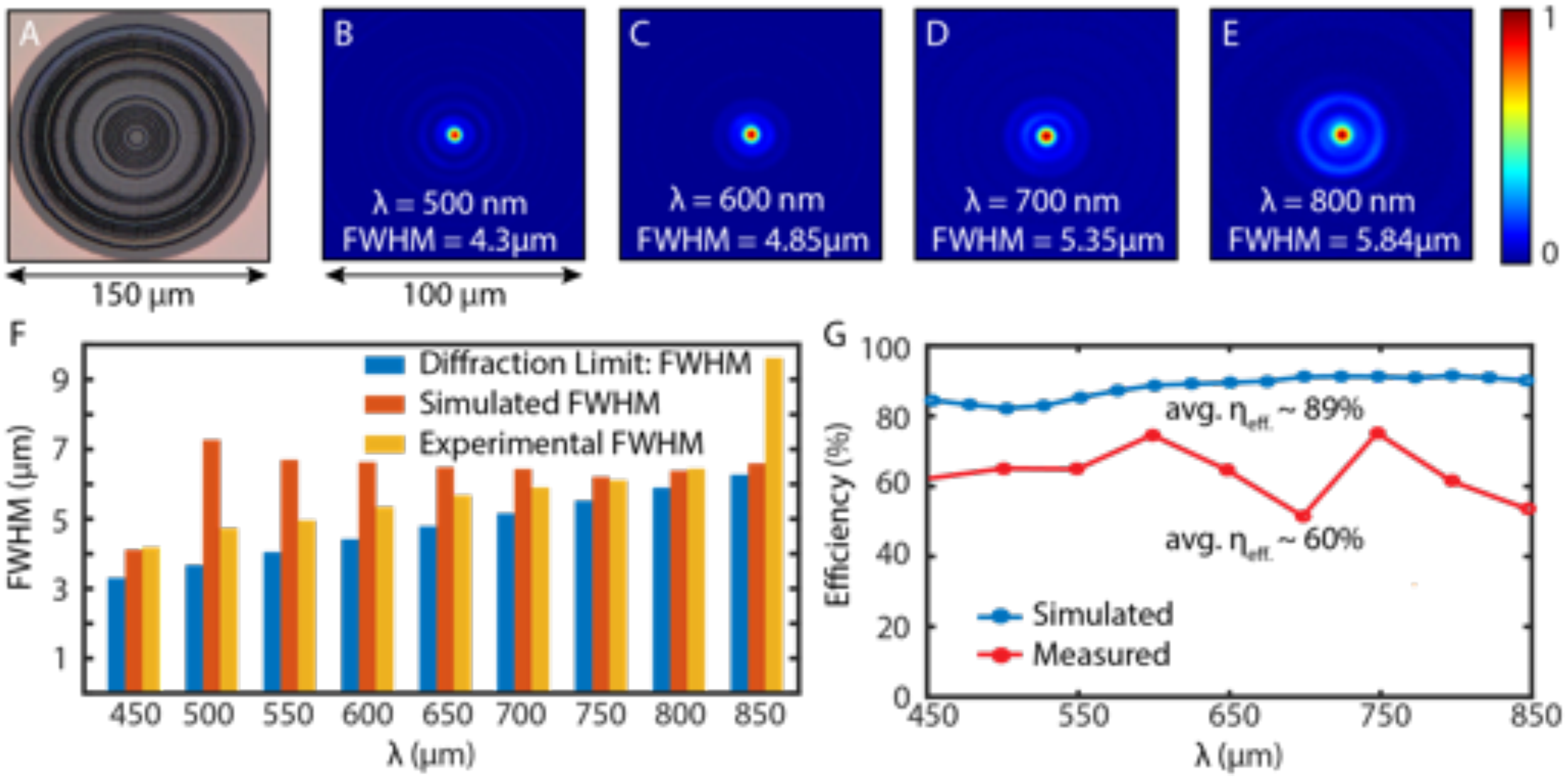

***Figure 2:*** *Achromatic focusing by the MDL. (A) Optical micrograph of the fabricated MDL. (B-E) Measured point-spread functions at the same focal plane, 1mm away from the MDL at four wavelengths. (F) Measured, simulated and diffraction-limited full-width at half-maximum (FWHM) as function of wavelength. (G) Measured and simulated focusing efficiency as a function of wavelength. The operating bandwidth of the MDL is 450nm to 850nm.*

The focusing efficiency of the MDL (Fig. 2G) is calculated as the power within a spot of diameter equal to 3 times the FWHM divided by the total power incident on the lens [16, 24]. The simulated wavelength averaged focusing efficiency for the designed MDL is ~89% in comparison to a measured value of ~60% in the 450 nm to 850 nm band. In order to explain this discrepancy, we performed careful simulations of the sensitivity of the focusing efficiency to errors in the ring heights (Fig. S8) and ring widths (Fig. S9) [16]. The analysis suggests that standard deviation in ring heights of 100nm coupled with standard deviation in ring widths of about 150nm can explain the drop in efficiency (Fig. S11). Such standard deviations are expected in our existing grayscale lithography process (see Fig. S10 for error measurements). In the future, it is possible to incorporate tolerance to fabrication errors as one of the metrics during the optimization-based design step, analogous to what was done previously for binary multi-wavelength diffractive lenses [25].

Next, we assembled a camera by placing the MDL in front of a conventional image sensor (Fig. 3A). Then, we characterized the imaging behavior of the MDL by capturing still and video images of the Air Force resolution target (450nm to 850nm in Fig. 3B and illumination spectrum in Fig. S18) and the Macbeth color chart (visible in Fig. 3C and in Supplementary Video 1, and NIR in Fig. 3D and in Supplementary Video 2) [16]. The distance between the MDL and the image sensor was ~2.5mm, and the distance between the object and the MDL was ~1.7mm. The visible and NIR frames were captured using a color image sensor (DFM 72BUC02-ML, Imaging Source). In each case, the exposure time was adjusted to ensure that the frames were not saturated. In addition, a dark frame was recorded and subtracted from the images. The resolution-chart image shows that the resolved spatial frequency is ~144 line-pairs/mm, which corresponds to a spatial period of 6.9μm or about 3 times the sensor pixel size (2.2μm). This resolution corresponds approximately to the average FWHM over all the wavelengths in Fig. 2F. Finally, the visible image of the Macbeth color chart shows excellent color reproduction.

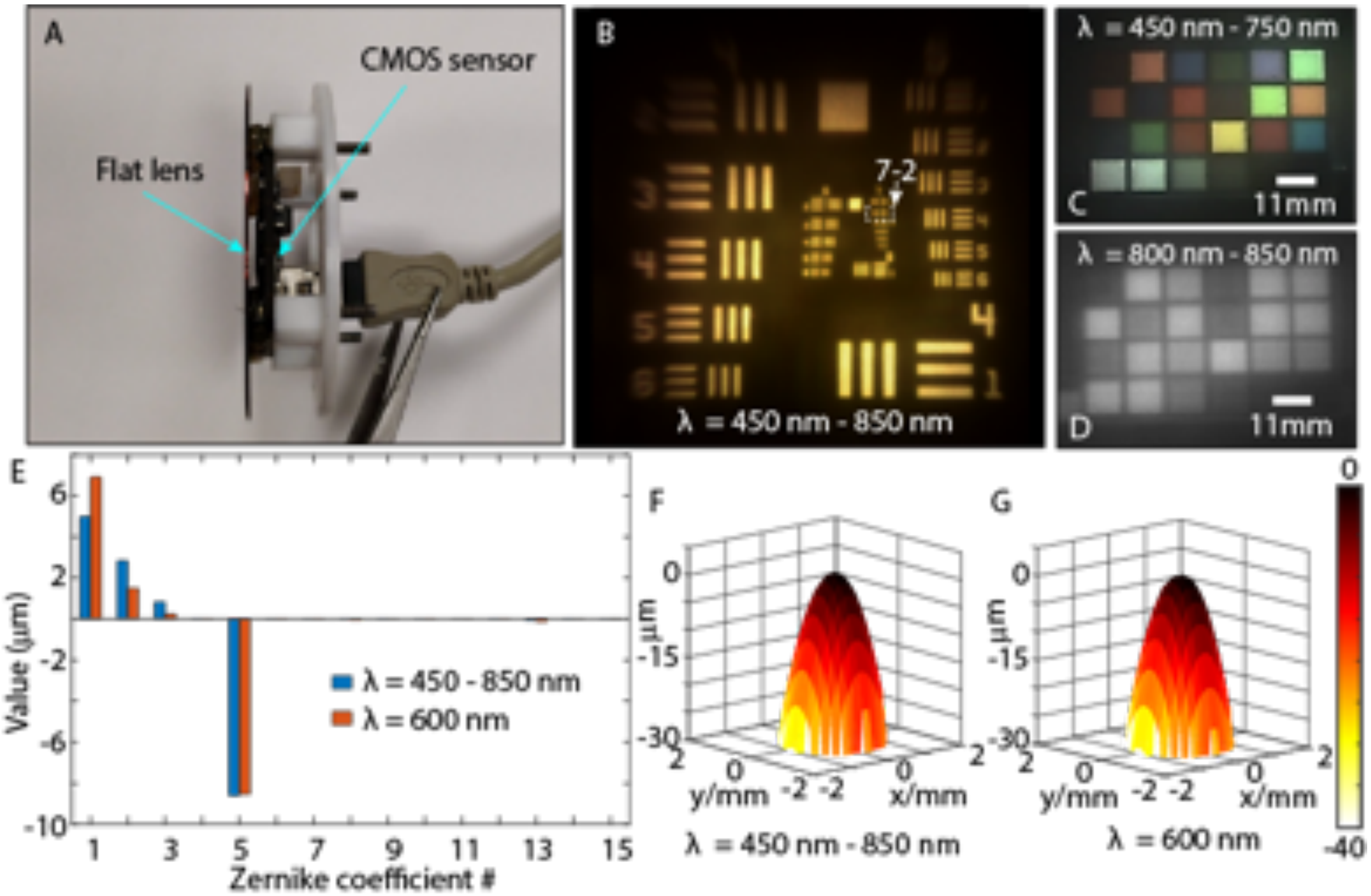

***Figure 3:*** *Imaging characterization of the MDL. (A) Photograph of the camera comprising a conventional image sensor and the MDL. (B) Image of the AirForce resolution chart under broadband (450nm to 850nm) illumination. Images of the Macbeth color chart under (C) visible (see Supplementary Video 1) and (D) near-IR illumination (see Supplementary Video 2). (E) Zernike aberrations coefficients measured using a Shack-Hartmann wavefront sensor for our MDL under broadband illumination and under 600nm*

*illumination. Measured wavefront errors under (F) broadband and (G) narrowband illumination. Note that the distance between the MDL and the image sensor was fixed while taking the visible and NIR images.*

The wavefront aberrations of the fabricated MDL were measured using a Shack-Hartmann wavefront sensor (Thorlabs, WFS 150-7AR) [16]. The wavefront aberrations were measured under broadband (450nm to 850nm) and under narrowband illuminations at 6000nm with 50 nm bandwidth. The corresponding Zernike polynomial coefficients are shown in Fig. 3E. The measurements confirm that the MDL indeed has low values for all aberrations (Table S2) [16]. Most importantly, the aberrations as well as the reconstructed wavefront are quite similar for both the broadband (Fig. 3F) and 600nm narrowband (Fig. 3G) illumination, confirming excellent achromaticity. We also simulated the aberrations for the MDL and these agree well with the experiments (Tables S3 and S4) [16].

Next, we calculated the modulation-transfer function (MTF) of the MDL by taking the absolute value of the Fourier transform of the captured PSFs. The extracted MTF plots are shown in Fig. 4A for all the wavelengths. The first zero crossing of the on-axis MTF at 10% contrast occurs between 130 to 140 line-pairs/mm for the visible wavelengths and about 60 line-pairs/mm for the NIR wavelengths. This is also quite consistent with the resolution limits estimated from the chart in Fig. 3. We furthered studied vignetting of the MDL (Fig. 4B) by photographing a uniform plane of white color, which was bigger than the field of view (FOV) of the MDL. This uniform white plane was illuminated from far by white LEDs. The normalized intensity across the MDL FOV with respect to its viewing angle gives us the vignetting performance of the MDL. For characterizing the geometric distortion of the MDL, we replaced the white plane with a regular checkered grid and measured the grid-point distortion percentage with respect to the MDL viewing angle. The results of the geometric distortion are shown in Figs. 4C and 4D in the vertical and horizontal directions, respectively. Low astigmatism is noted. Lastly, we measured the signal-to-noise ratio (SNR) response of the MDL when photographing uniform planes of red, green and blue colors (Figs. E-G, respectively) and confirm SNR > 40dB for all corresponding color channels. Our careful characterization shows that the performance of our MDL is comparable to that of thick refractive lens systems.

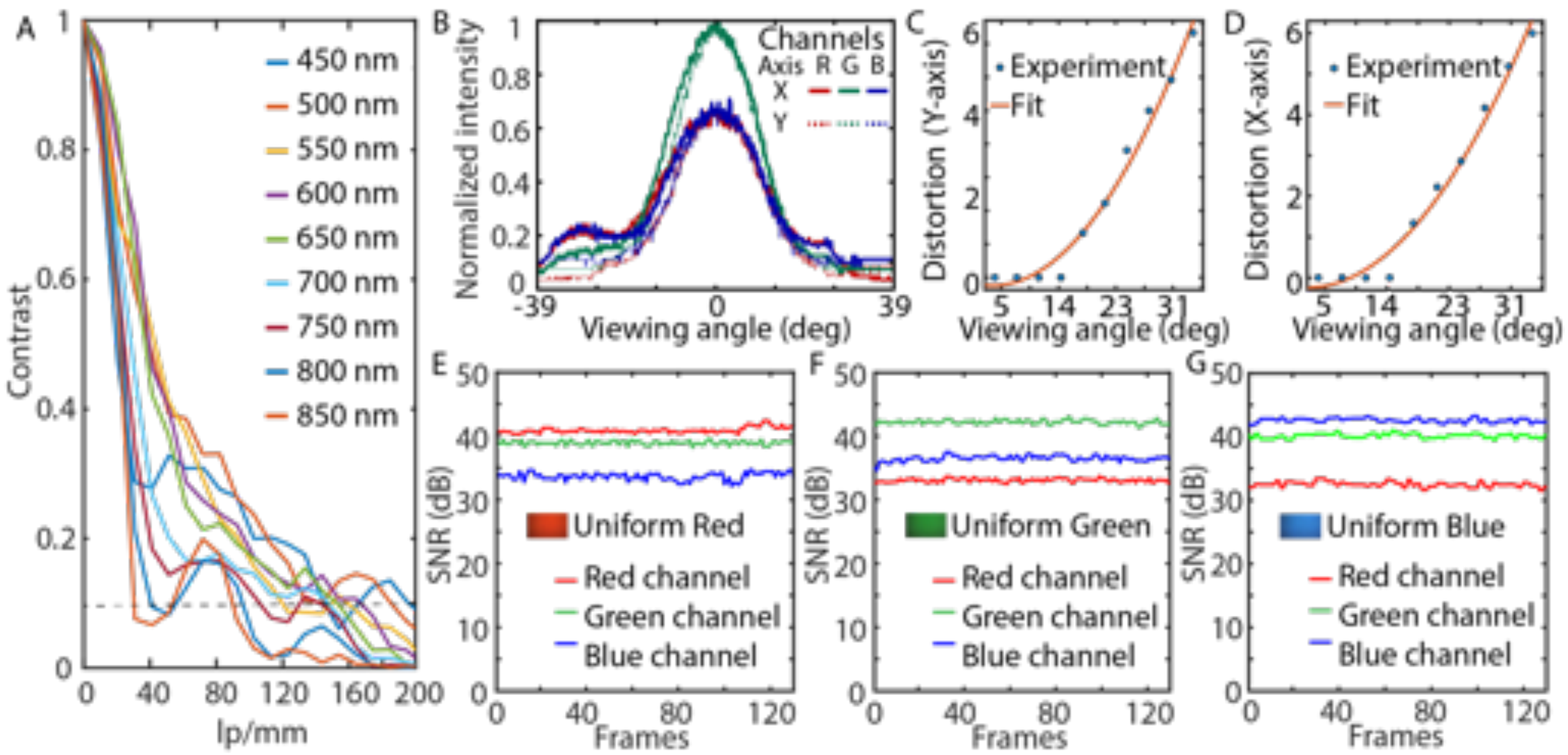

***Figure 4**: Analysis of the MDL. (A) Modulation-transfer-function (MTF) and (B) Vignetting measurement showing normalized intensity vs viewing angle in degrees. (C, D) Geometric distortion showing distortion of a regular geometric grid in the Y (C) and X (D) axes as a function of the viewing angle in degrees and (E-G) Signal-to-Noise Ratio (SNR) of the MDL when imaging a patch of uniform (E) Red, (F) Green and (G) Blue color for each of the Red, Green and Blue channels of the CMOS sensor.*

**Extending the operating bandwidth**

Extending the concept of phase in the focal plane as a free parameter, we can design a single MDL (one diffractive surface) to be achromatic over an even larger bandwidth. In Fig. 5A, we show the simulated focusing efficiency of a lens designed to focus $\lambda$=0.5μm to 15μm at a single focal plane, 5mm away. The dispersion properties of photopolymer (AZ9260, Microchem) (Fig. S2) were assumed [16]. We have previously fabricated MDLs in the IR using this material [21]. Even though the photopolymer exhibits absorption in the IR, due to its drastically reduced thickness (10μm), the lens exhibits focusing efficiency of 76% averaged over the entire wavelength range of 0.5μm to 15μm. The designed geometry of the lens is shown in Fig. 5B. The minimum feature width, maximum feature height and number of levels were 6μm, 10μm and 100, respectively. The simulated point-spread functions at select wavelengths are shown in Figs. 5C-H (PSFs for all the wavelengths are shown in Fig. S19 [16]). We further confirmed that the simulated full-width at half-maximum approximately tracked the theoretical diffraction-limited values (Fig. S20) [16]. In addition, we also performed a rigorous finite-difference time domain (FDTD) simulation of this MDL for select wavelengths to highlight the achromatic focusing in a 2D slice (Fig. S21) [16].

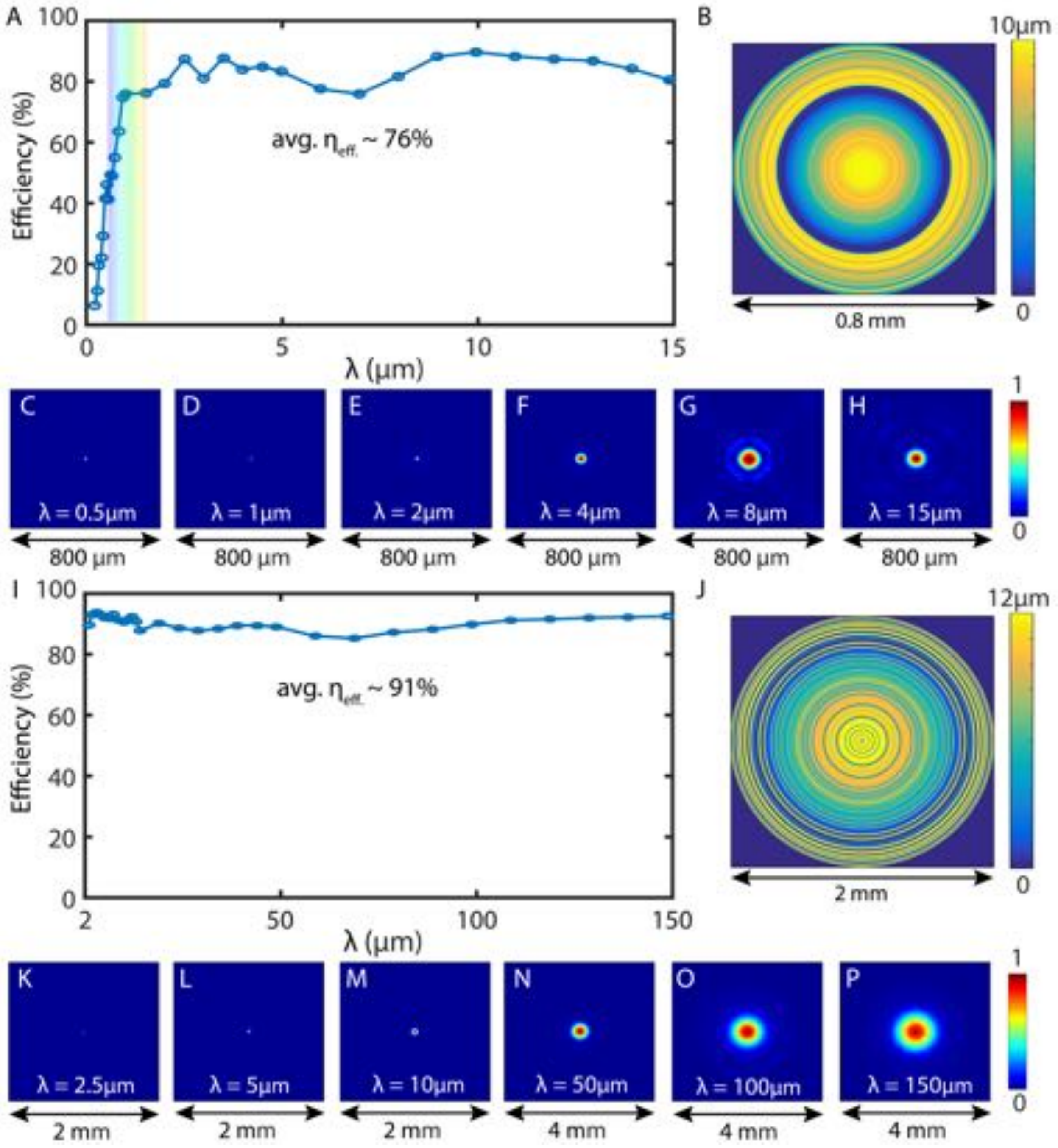

***Figure 5**: Achieving unlimited bandwidths. (A) Focusing-efficiency spectrum of an MDL with f=5mm and aperture=0.8mm for λ =0.5μm to 15μm assuming a photopolymer material. (B) Geometry of the MDL. The minimum feature width, maximum feature height and number of levels were 6μm, 10μm and 100, respectively. (C-H) Simulated PSFs at exemplary wavelengths spanning the bandwidth. (I) Focusing-efficiency spectrum of an MDL with f=10mm and aperture of 2mm for λ=2.5μm to 150μm assuming Silicon as the MDL material. (J) MDL geometry. The minimum feature width, maximum feature height and number*

*of levels were 6μm, 12μm and 100, respectively. (K-P) Simulated PSFs at exemplary wavelengths spanning the bandwidth.*

We expect the performance of the lens to be directly related to the number of degrees of design freedom, which is the product of the number of rings in the radially symmetric lens and the number of height levels that can be occupied by each ring. We explored the impact of these degrees of freedom via simulations (Figs. S22 and S23) [16]. The average focusing efficiency of the MDL decreases with decreasing number of degrees of freedom as expected. However, the dependence seems to be relatively weak. Therefore, we attempted to design an MDL with an even larger operating bandwidth of λ=2.5μm to 150μm. Note that we chose this bandwidth only because we were able to find the dispersion data for a transparent material (Silicon) for these wavelengths in literature (Fig. S3) [16]. The focusing efficiency of this MDL is shown as a function of wavelength in Fig. 5I. Note that the focusing efficiency is 91% averaged over λ=2.5μm to 150μm. We remind the reader that this is achieved with a single diffractive surface! The height distribution of this MDL is shown in Fig. 5J. The minimum feature width, maximum feature height and number of levels were 6μm, 12μm and 100, respectively. Exemplary simulated point-spread functions are shown in Figs. 5K-P (the corresponding plots for all the wavelengths are in Fig. S24 [16]). We also confirmed that the simulated FWHM closely tracks the diffraction limit (Fig. S25) [16]. Clearly, the MDL is achromatic over the entire wavelength range from 2.5μm to 150μm. In order to confirm that both MDLs will be useful for imaging, we simulated the wavefront aberrations as a function of wavelength and confirmed that the aberrations are indeed minimal across all the design wavelengths (Tables S5 and S6) [16]. We note again that we do not know the limit of this operating bandwidth. Our conjecture is that the bandwidth will be limited by the channel capacity theory dictated by Information Theory [26]. Nevertheless, allowing phase to be a free parameter seems to drastically reduce the constraints enabling a far larger operating bandwidth than we originally anticipated.

**Conclusion**

We describe a generalized approach to lens design that allows the phase distribution in the image (or focal) plane to be a free parameter. This freedom allows the lens transmittance function to have numerous solutions, and we propose that the choice of solution could be made to enable achromaticity, manufacturability, minimization of aberrations, of weight, of thickness, of cost, etc. We utilize this concept to design a single diffractive surface that is achromatic with a huge bandwidth (as large as 2.5μm to 150μm) with no apparent practical limits. Our approach can be readily generalized to metasurfaces (by employing full-wave diffraction models), which could be advantageous to manipulate the polarization states of light [27] and could even be implemented in integrated-photonics platforms [28-30].

## Methods

### Design and Optimization

The multi-level diffractive lens (MDL) is modeled using scalar diffraction theory in the regime of Fresnel approximation. The beam propagation is modelled with the Fresnel-Kirchhoff diffraction integral as given in eqn. (1):

$$U(x',y',\lambda,d)=\frac{e^{ikd}}{i\lambda d}\iint g_{illum}(x',y',\lambda)\ .T(x',y',\lambda).e^{i\frac{k}{2d}[(x-x')^2+(y-y')^2]}dx \qquad (1)$$

where $T(x,y,\lambda)$ is the transmission function at the lens surface and $g_{illum}(x',y',\lambda)=1$ is the on-axis unit amplitude illumination wave. The intensity at the focal plane is given by $I(x',y',\lambda,d)=|U(x',y',\lambda,d)|^2$. We utilized a modified version of direct-binary search to optimize the height profile of rotationally symmetric MDLs consisting of individual constituent rings of width equal to a pre-defined value.

### FDTD Simulation

For the 0.5um to 15um broadband MDL design, an additional full wave 2D FDTD simulation was done to verify the achromaticity of the design as predicted from the scalar prediction. A full 3D simulation could not be pursued due to extreme computational requirement both in terms of memory and time complexity. Symmetry along the x-axis is exploited to reduce the simulation time by 1/2 of the original simulation requirements. Moreover, even with 2D FDTD simulation wavelengths smaller than 5um is again computationally expensive and has been avoided. Full details of the FDTD simulation setup has been provided in the supplementary information.

**Fabrication and Characterization**

The achromatic diffractive lenses were patterned on a photoresist (S1813) film atop a glass wafer using grayscale laser patterning using a Heidelberg Instruments MicroPG101 tool. The exposure dose was varied as a function of position in order to achieve the multiple height levels dictated by the design. After fabrication, the devices were characterized on an optical bench by illuminating them with broadband collimated light, whose spectral bandwidth could be controlled by a tunable filter. The focus of the lenses were captured on a monochrome CMOS sensor for characterization of the PSF. Imaging performance of the lenses were tested in prototype cameras as described in the main text with various objects namely the Macbeth chart and USAF resolution chart.

**Acknowledgements**

We thank Brian Baker, Steve Pritchett and Christian Bach for fabrication advice, and Tom Tiwald (Woollam) for measuring dispersion of materials. RM acknowledges useful discussion with Fernando Vasquez-Guevara. We would also like to acknowledge a grant of credit on Amazon AWS (051241749381) for computation. The support and resources from the Center for High Performance Computing at the University of Utah are also gratefully acknowledged. RM and BSR acknowledges funding from the Office of Naval Research grant N66001-10-1-4065 and from an NSF CAREER award: ECCS #1351389, respectively.

**Competing Interests Statement**

RM is co-founder of Oblate Optics, Inc., which is commercializing technology discussed in this manuscript. The University of Utah has filed for patent protection for technology discussed in this manuscript.

## Author Contributions

RM, BSR and SB conceived and designed the experiments. SB and RM modeled and optimized the devices. MM fabricated the devices. SB and MM performed the PSF experiments. MM, SB and AM performed the imaging experiments. SB performed the MTF analysis. SB and AM performed the aberrations analysis. AM performed the distortion analysis. All authors performed the data analysis and wrote the manuscript.

## Materials and Correspondence

Correspondence and materials requests should be addressed to RM at rmenon@eng.utah.edu.

**Supplementary Information**

# Imaging over an unlimited bandwidth with a single diffractive surface

*Sourangsu Banerji,* [1,*] *Monjurul Meem,* [1,*] *Apratim Majumder,* [1] *Berardi Sensale-Rodriguez*[1] *and Rajesh Menon*[1,2, a)]

[1]Department of Electrical and Computer Engineering, University of Utah, Salt Lake City, UT 84112, USA.

[2]Oblate Optics, Inc. San Diego CA 92130, USA.

a) rmenon@eng.utah.edu

[*] Equal contribution.

## 1. Summary of the Literature survey of broadband achromatic metalenses

**Table S1.** Broadband Metalenses

| **Material** | **Wavelength** | **Bandwidth** | **N.A.** | **Focal Length/ Diameter** | **Simulated / Measured Efficiency** | **Feature Width/Height** | **Polarization** | **Reference** |
|---|---|---|---|---|---|---|---|---|
| a-Si | 1.3 um - 1.65 um | 350 nm | 0.24 | 200 um / 100 um | average 35% (measured)- | 100 nm / 1400 nm | Polarization Insensitive | [1] |
| a-Si | 1.2 um - 1.65 um | 450 nm | 0.13 | 800 um / 200 um | average 32% (measured) | 100 nm / 800 nm | Polarization Insensitive | [1] |
| a-Si | 1.2 um - 1.40 um | 200 nm | 0.88 | 30 um / 100 um | - | 100 nm / 800 nm | Polarization Insensitive | [1] |
| GaN | 400 nm - 660 nm | 260 nm | 0.106 | 235 um/ 50 um | average 40% (measured) | 45 nm/800 nm | Circular | [2] |
| TiO2 | 460 nm - 700 nm | 240 nm | 0.2 | 67um /26.4um | average 35% (measured | 50 nm /600 nm | Polarization Insensitive | [11] |
| a-Si | 1300 nm - 1800 nm | 500 nm | 0.04 | 7.5 mm / 600 um | 24%, 22%, and 28% | 75 nm/ 600 nm | Linear | [3] |
| Au/SiO2/Au | 1.2 nm - 1.68 um | 480 nm | 0.268 | 100 um / 55.55 um | 12.44 % (measured) | 40nm /30 nm | Circular | [4] |
| Au/SiO2/Au | 1.2 nm - 1.68 um | 481 nm | 0.217 | - | 8.4 % (measured) | 40nm /30 nm | Circular | [4] |
| Au/SiO2/Au | 1.2 nm - 1.68 um | 482 nm | 0.324 | - | 8.56 % (measured) | 40nm /30 nm | Circular | [4] |
| TiO2 | 470 nm - 670 nm | 200 | 0.2 | 63 um / 25.2 um | 50% (Simulated) | 80 nm / 600 nm | Circular | [5] |
| PbTe | 5.11 um - 5.29 um | 180 nm | 0.5 mm / - | - | - | / 650 nm | - | [12] |
| GaSb | 3 um - 5 um | 2 um | 0.35 | 155 um / >300 um | 70% (Simulated) | 30 um / 2 um | Polarization Insensitive | [13] |
| a-Si | 5 um - 8 um | 3 um | 0.35 | 30*lambda / - | - | - / 1.5*lambda | Polarization Insensitive | [6] |
| a-Si | 3.7 um - 4.2 um | 0.5 um | 0.45 | 300 um / 300 um | ~ 96% (Simulated) | / 2 um | Polarization Insensitive | [15] |
| GaN | 435 nm - 685 nm | 250 nm | 0.17 | 20 um / 7 um | 50% - 78% (Simulated) | 160nm or 240 nm / 400 nm | Linear | [7] |
| Photoresist ( polymer-ZEP520A) | 436 nm - 685 nm | 250 nm | 0.17 | 20 um / 7 um | 50% - 78% (Simulated) | 160nm or 240 nm / 400 nm | Linear | [7] |
| PbTe | 5.11 um - 5.29 um | 180 nm | - | 0.5 mm / 1 mm | ~ 75% (Simulated) | 2500 nm / 650 nm | Linear | [16] |
| Au | 532 nm - 1080 nm | 548 nm | - | 7 um / 10 um | ~ 20% (measured) | 100 nm , 60 nm, 40 nm / 40 nm | Linear | [8] |
| a-Si | 470 nm - 658 nm | 188 nm | 0.3511 | 400 um / 300 um | Qualitative Agreement | - / 15 to 50 nm | Polarization Insensitive | [14] |
| TiO2 | 490 nm - 550 nm | 60 nm | 0.2 | 485 um / 200 um | Qualitative Agreement | varied / 180 nm | - | [9] |
| TiO2 | 560 nm - 800 nm | 240 nm | upto 0.8 | 2 um to 14 um / 5.4 um | > 50% (Simulated) | 10 nm - 150 nm / 488 nm | Polarization Insensitive | [17] |
| fused Si | 486 nm - 656 nm | 170 nm | 0.1 | 100 mm / 20 mm | - | 1300 nm / 560 nm | Polarization Insensitive | [10] |

This discrepancy in the definition of focusing efficiency is observed across all the papers on metalens design. The definition of focusing efficiency in the highlighted broadband metalens papers above in Table S1 are summarized below:

**Reference 1:**
**Definition:**
To measure the focusing efficiency, the motor is first moved to the focal plane of the metalens and the iris is closed such that it corresponds to three to five times the FWHM of the focal spot. Then, the light is passed to the power meter. The ratio between the measured focused power and the incident power defines the focusing efficiency.

**Note:** The paper states, "the maximum shift from the mean focal length limited to 2–5% for the entire design bandwidth." It is not clear from the paper whether the authors took 3 times or 5 time the FWHM of the focal spot.

**Reference 2:**
**Definition:**
The efficiency is defined as the ratio of the optical power of the focused circularly polarized beam to the optical power of the incident beam with opposite helicity.

**Reference 3:**
**Definition:**
Absolute efficiency is defined as the amount of power in the beam waist at the focal line, divided by the input power. The beam waist is the full width of the normalized intensity at $1/e^2$ and is calculated by fitting the measured intensity distribution at the focal line to a Gaussian function. The input power is the amount of collimated power (beam diameter of 4 mm) that passes through an aperture with the same dimensions as the lens (600-μm × 600-μm).

**Reference 4:**
**Definition:**
The efficiency is defined as the ratio of light intensity from the focal spot at corresponding focal plane to the light intensity reflected by a metallic mirror with the same pixel sizes. (For circularly polarized incidences in a reflection scheme)

**Reference 8:**
**Definition:**
The transmission efficiency is defined as the ratio of power transmitted through the lattice over the incident intensity. The diffraction efficiency, or the power focused within the focal point, was between 1.5% and 6%.

**Reference 11:**
**Definition:**
The achromatic metalens was also characterized by measuring the focusing efficiency of the focal spot under different polarizations of incident light. The focusing efficiency is defined as the focal spot power divided by transmitted power through an aperture with the same diameter as the metalens. The measured focusing efficiencies weakly change with polarization.

**Note**: The paper states "The metalens exhibits a measured focal length shift of only 9% λ = 460–700 nm and has diffraction-limited focal spots across the entire range. The focusing efficiency of the metalens varies by only ~ 4% under various incident polarizations."

## 2. Material dispersion

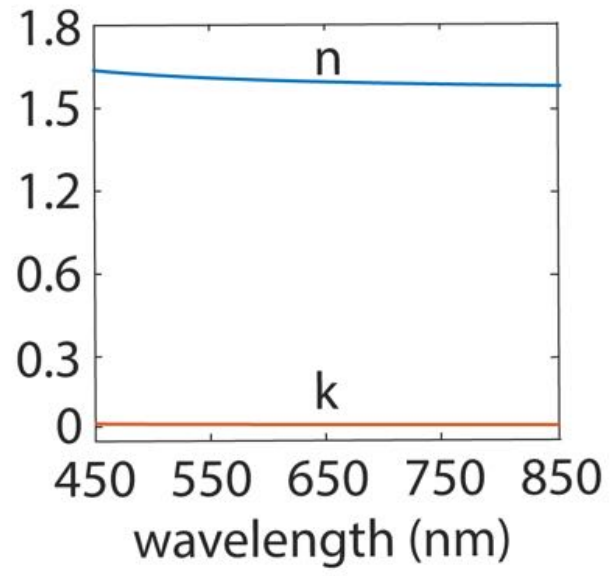

***Fig. S1:*** *Dispersion of S1813. This is used for the lens in Figs. 1-4.*

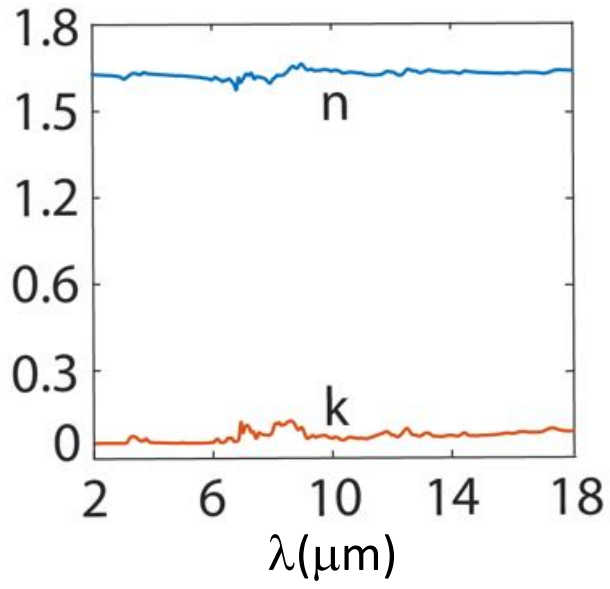

***Fig. S2:*** *Dispersion of AZ9260. This is used for the lens in Figs. 5A-H.*

***Note:*** *A value of n = 1.61 and k = 0 was assumed for all wavelengths below 2 μm (typically the visible band) during optimization.*

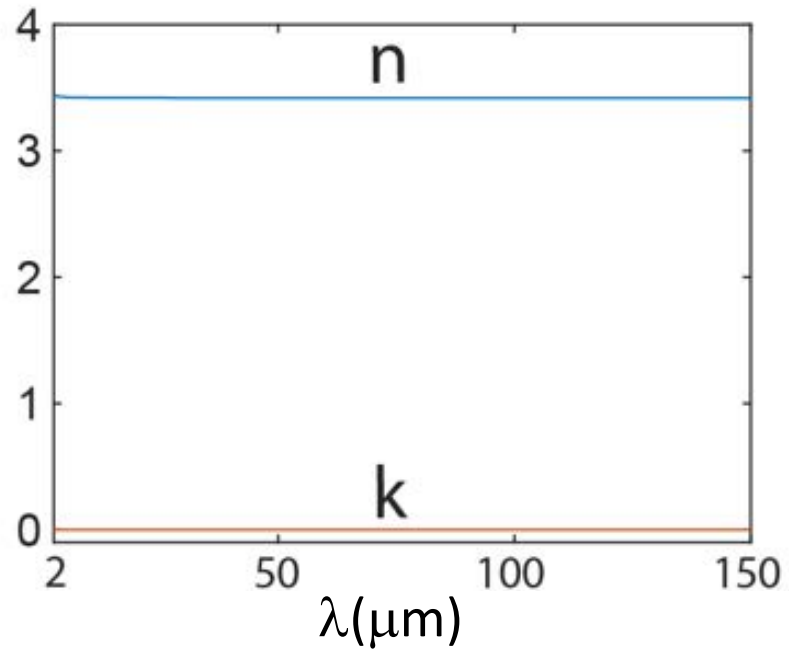

***Fig. S3:*** *Dispersion of silicon [26,27]. This is used for the lens in Figs. 5I-P.*

## 3. Simulated and Measured Point Spread Functions (PSFs) of the 450nm – 850nm broadband MDL

The scalar simulated point spread functions (PSFs) for the broadband lens designed are provided below:

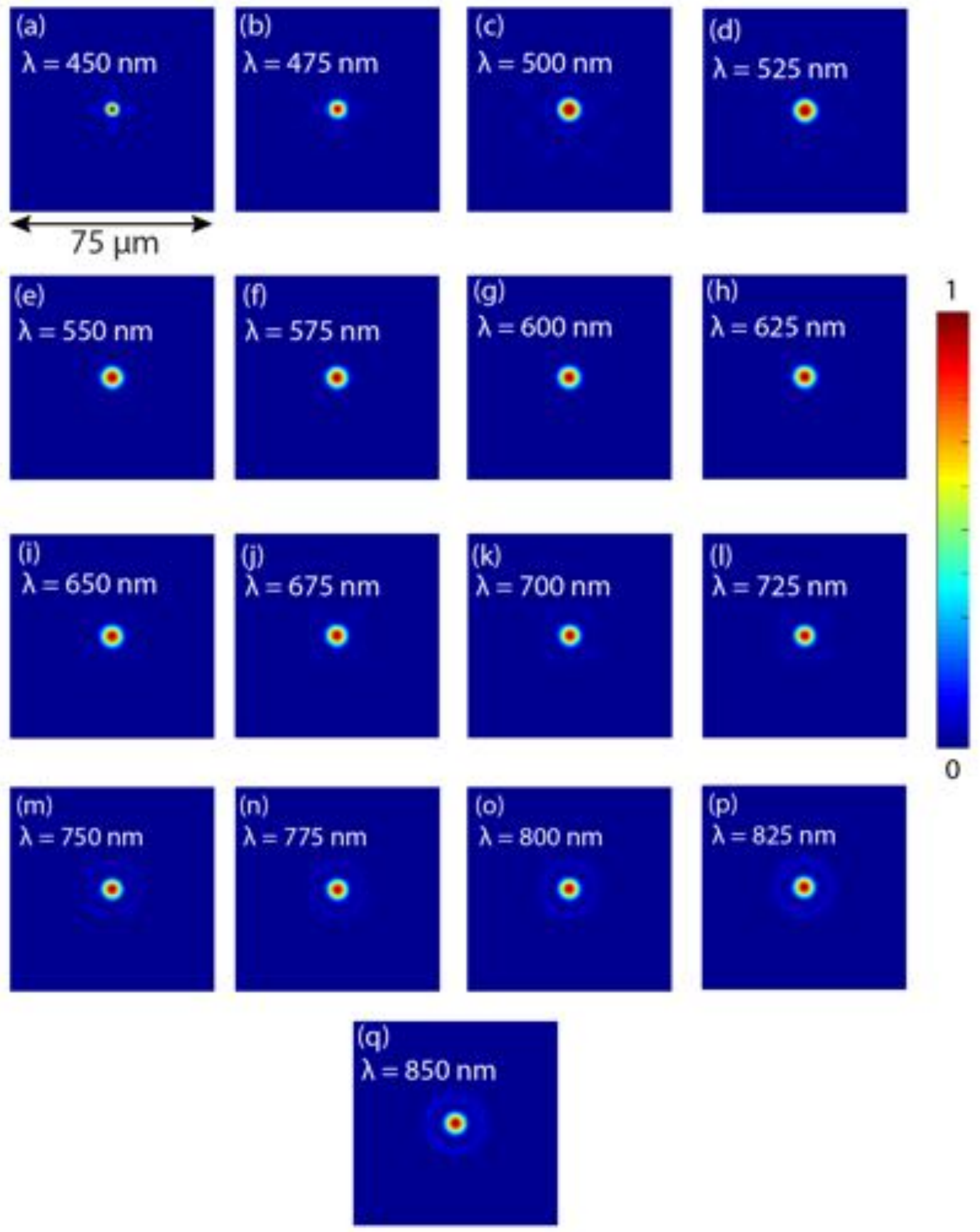

***Fig. S4:*** *Simulated PSFs for* ***(a)*** *450nm* ***(b)*** *475 nm* ***(c)*** *500 nm* ***(d)*** *525 nm* ***(e)*** *550 nm* ***(f)*** *575 nm* ***(g)*** *600 nm* ***(h)*** *625 nm* ***(i)*** *650 nm* ***(j)*** *675 nm* ***(k)*** *700 nm* ***(l)*** *725 nm* ***(m)*** *750 nm* ***(n)*** *775 nm* ***(o)*** *800 nm* ***(p)*** *825 nm and* ***(q)*** *850 nm.*

The measured point spread functions (PSFs) for the broadband lens designed are provided below:

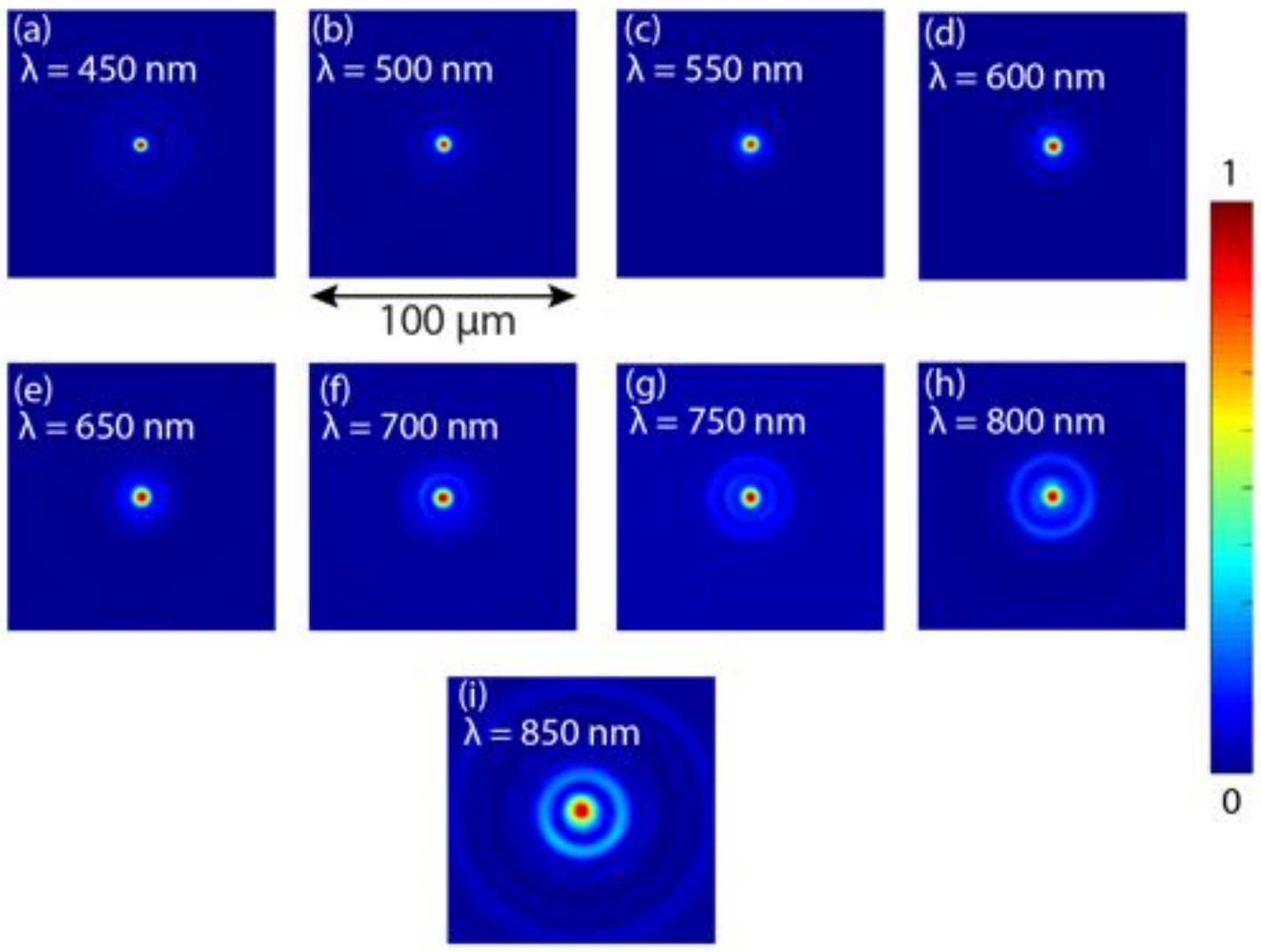

***Fig. S5:*** *Measured PSFs for* ***(a)*** *450nm* ***(b)*** *500 nm* ***(c)*** *550 nm* ***(d)*** *600 nm* ***(e)*** *650 nm* ***(f)*** *700 nm* ***(g)*** *750 nm* ***(h)*** *800 nm and* ***(i)*** *850 nm.*

**4. Fabrication**

The achromatic MDL depicted was patterned in a photoresist (Microchem, S1813) film atop a glass wafer (thickness ~ 0.6 mm) using grayscale laser patterning with a Heidelberg Instruments MicroPG101 tool [18-20]. In such conventional gray-scale lithography, the write head scans through the sample surface and the exposure dose at each point is modulated with different gray-scales [18, 19] (see schematic illustration of Fig. S4). Most of these typical photoresists are characterized by a contrast curve. Different depths in accord with different exposure doses are achieved after development. Greater dose leads to deeper feature. Before patterning structures, it is needed to calibrate this contrast curve. In this case, too, the exposure dose was varied with respect to position to achieve the multiple height levels dictated by the design.

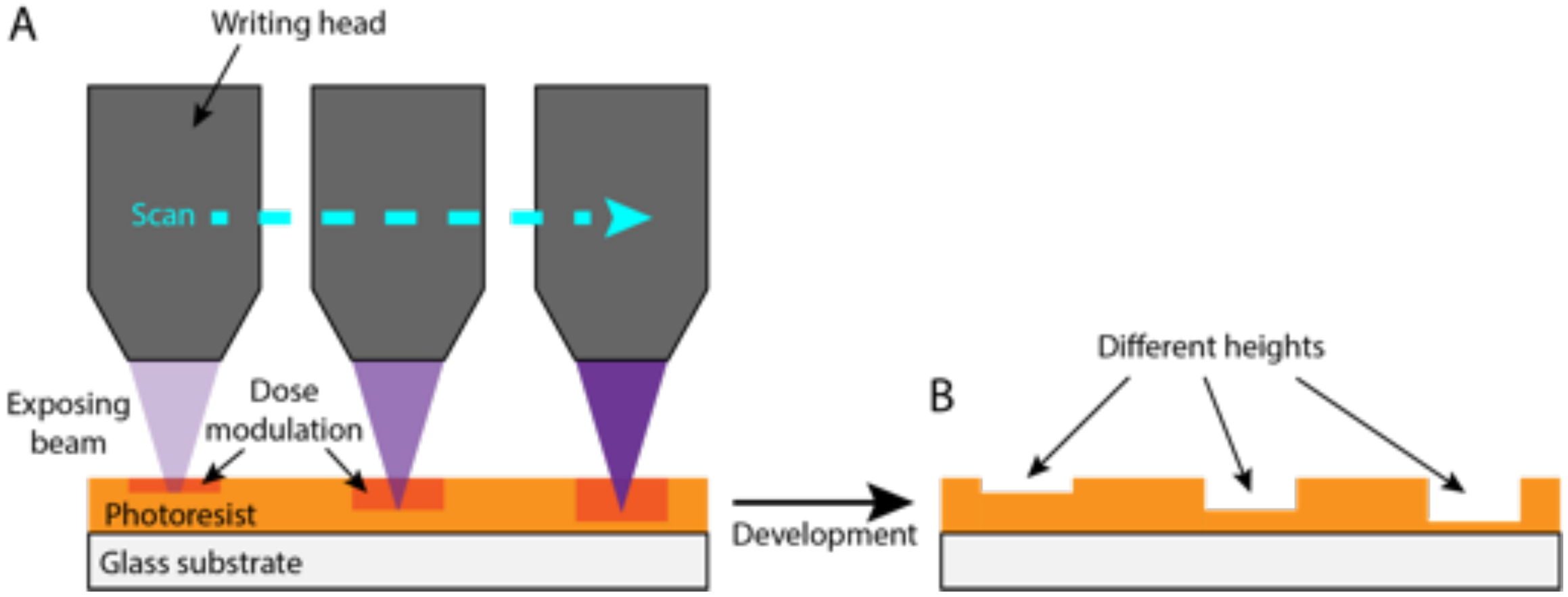

***Fig. S6:*** *Schematic of grayscale lithography used for fabricating the MDLs.*

**5. Experiment details**

**5.1. Focal spot characterization**

The flat lenses were illuminated with expanded and collimated beam from a SuperK EXTREME EXW-6 source (NKT Photonics) and the SuperK VARIA filter (NKT Photonics). The wavelength and bandwidth can be changed using the VARIA filter. The focal planes of the flat lenses were magnified using an objective (RMS20X-PF, Thorlabs) and tube lens (ITL200, Thorlabs) and imaged onto monochrome sensor (DMM 27UP031-ML, Imaging Source). The gap between objective and tube lens was ~90 mm and that between the sensor and the backside of tube lens was about 148mm. The magnification of the objective-tube lens was 22.22X.

After capturing the focal spot, the Focusing efficiency was then calculated using the following equation: Focusing efficiency = (sum of pixel values in 3*FWHM) / (sum of pixel values in the entire lens area)

### 5.2. Image characterization

For resolution chart imaging, the 1951 USAF resolution test chart (R3L3S1N, Thorlabs) was used as the object. The flat lenses were used for imaging the object on to the sensor. A diffuser was placed behind the USAF target. The experimental setup is shown in Fig. S7. The USAF target was illuminated with the design wavelengths with 10nm bandwidth and corresponding images were captured using a monochrome sensor (DMM 27UP031-ML, Imaging Source). The exposure time was adjusted to ensure that the images did not get saturated. In each case, a dark frame was recorded and subtracted from the USAF target images.

For imaging colorful objects, the objects were placed in front of the MDL just like the resolution chart. But this time the objects were illuminated with while LED light instead of a laser and corresponding images were captured using a color sensor (DFM 72BUC02-ML, Imaging Source).

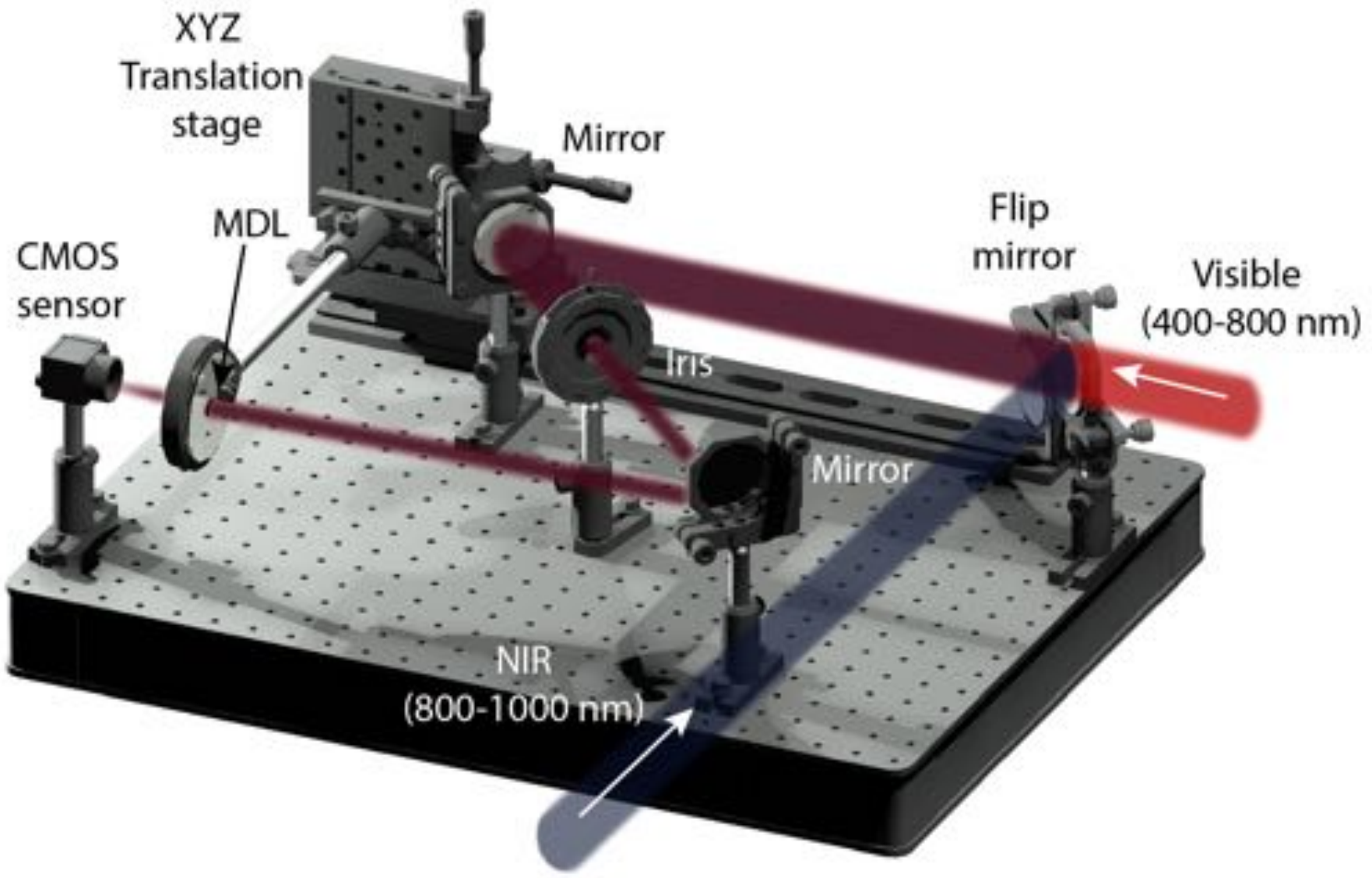

***Fig. S7***: *Schematic of system used for focusing experiments.*

The following supplementary videos are also included:

- Supplementary Video 1 shows a Macbeth color chart being moved around under visible light (450nm to 750nm).
- Supplementary Video 1 shows a Macbeth color chart being moved around under near-IR light (800nm to 850nm).

### 5.3. Simulated fabrication error analysis

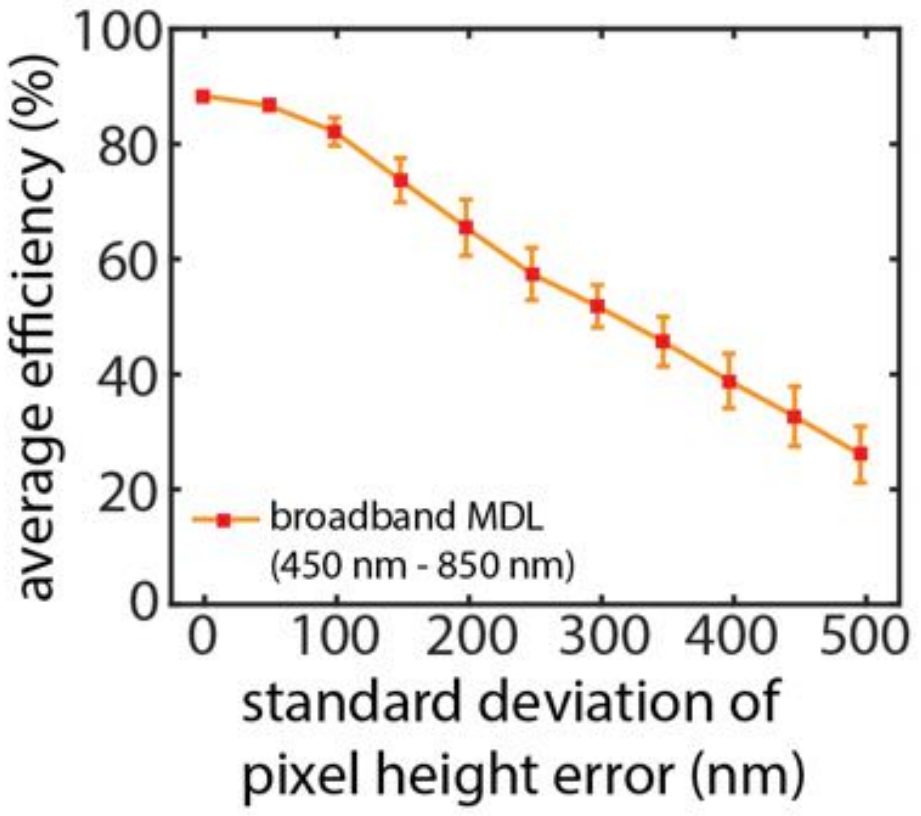

***Fig. S8:*** *Standard deviation based random pixel-height error analysis for the MDL designs.*

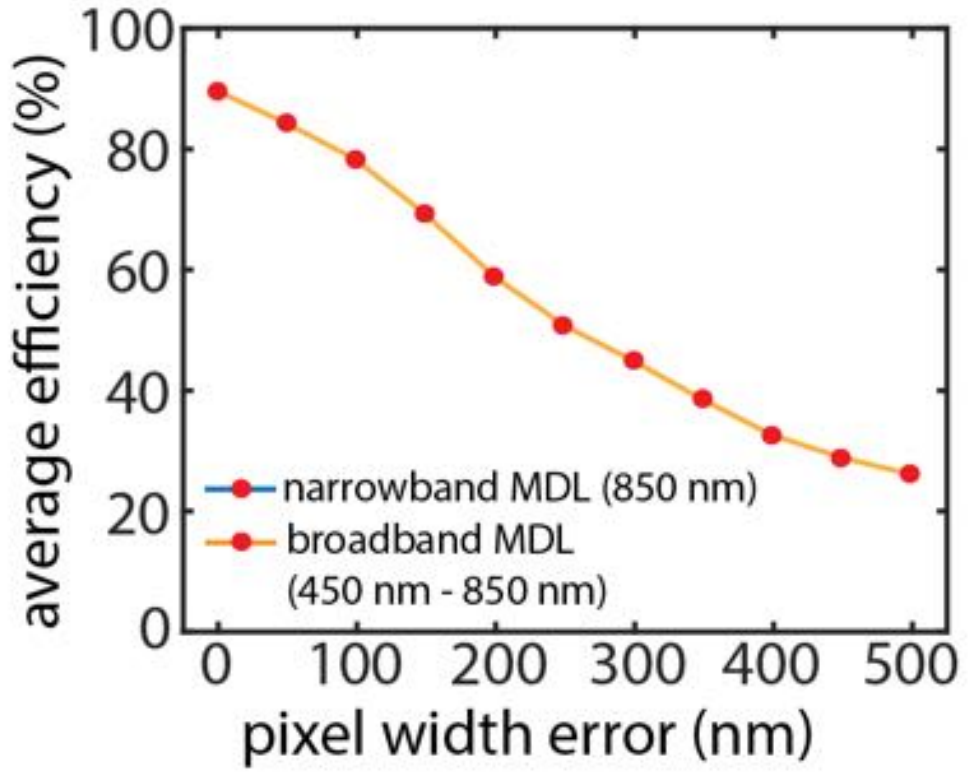

***Fig. S9:*** *pixel-width error analysis for the MDL designs*

We measured the pixel width of 20 randomly selected zones and these are summarized in Fig. S10 along with the corresponding design zone width. The estimated error has a mean of 270nm and standard deviation of 166nm.

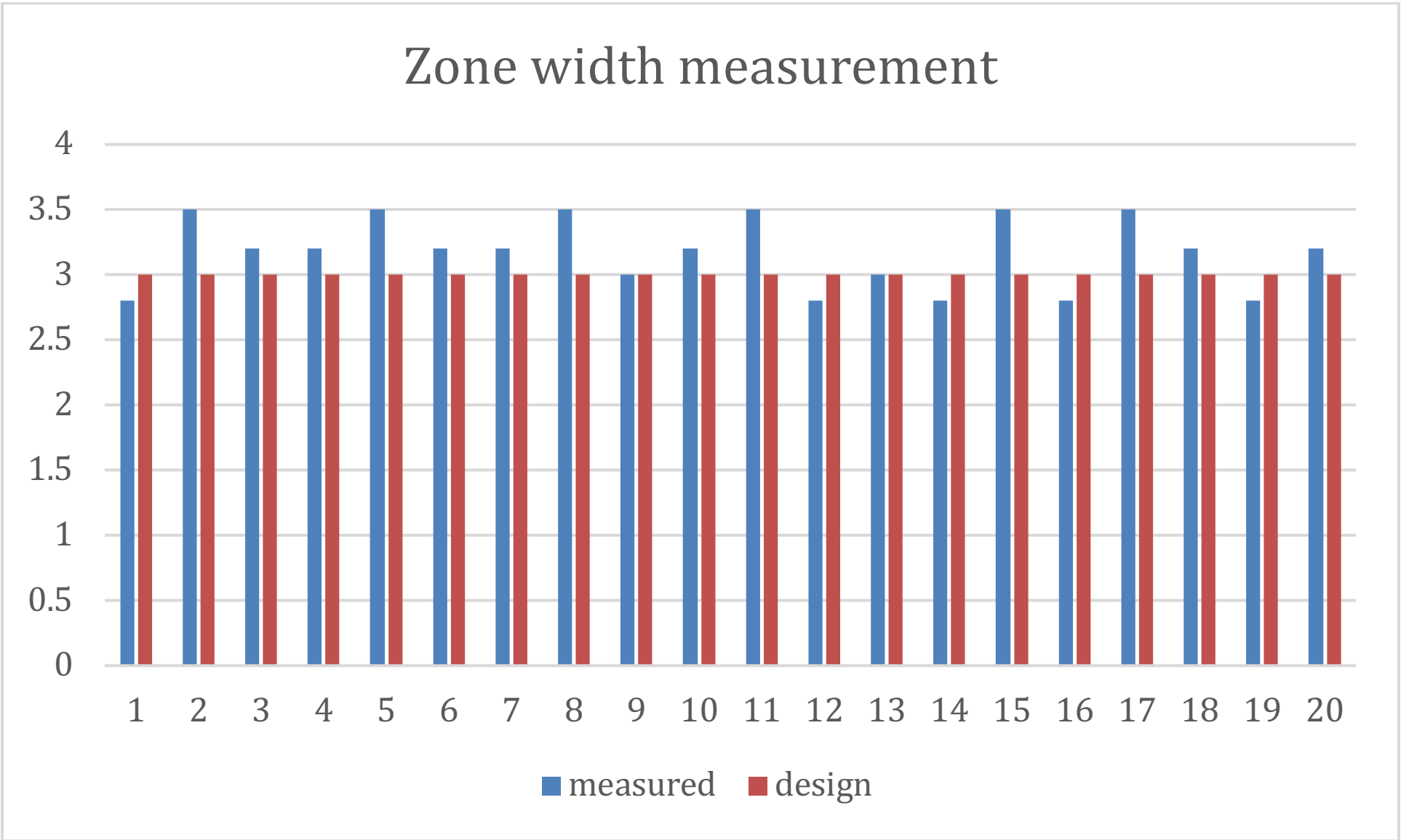

***Fig. S10:*** *Measured and design zone width of 20 randomly selected zones.*

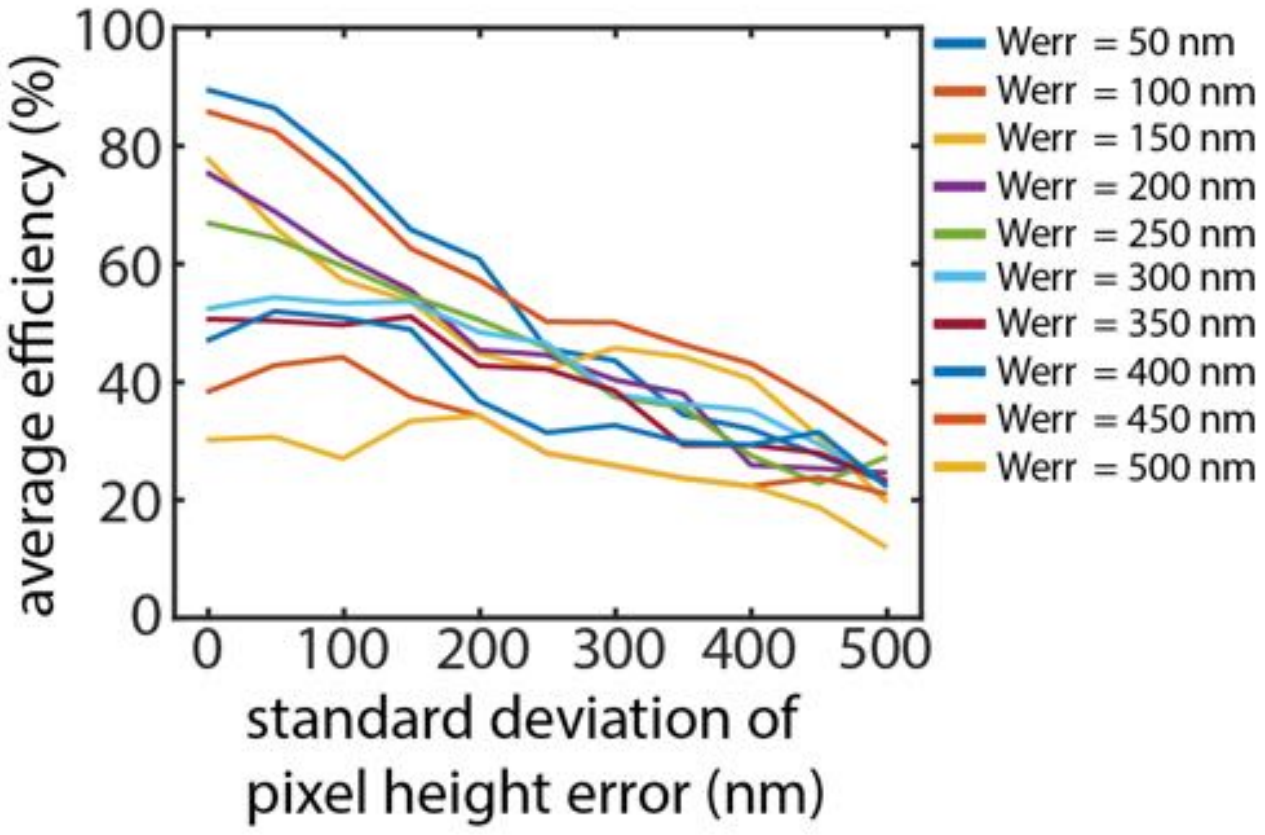

*Fig. S11:* *Pixel-width + standard deviation based pixel-height error analysis for the broadband MDL design (Werr denotes error in pixel-width)*

**5.4. Resolution from the USAF 1951 chart**

Resolution test targets are typically used to measure the resolution of an imaging system. They consist of reference line patterns with well-defined thicknesses and spacing, which being designed to be kept in the same plane as the object being imaged. By identifying the largest set of non-distinguishable lines, one determines the resolving power of a given system. The R3L3S1N from Thorlabs (as used here) negative target uses chrome coating to cover the substrate, leaving the pattern itself clear, and works well in back-lit and highly illuminated applications.

A

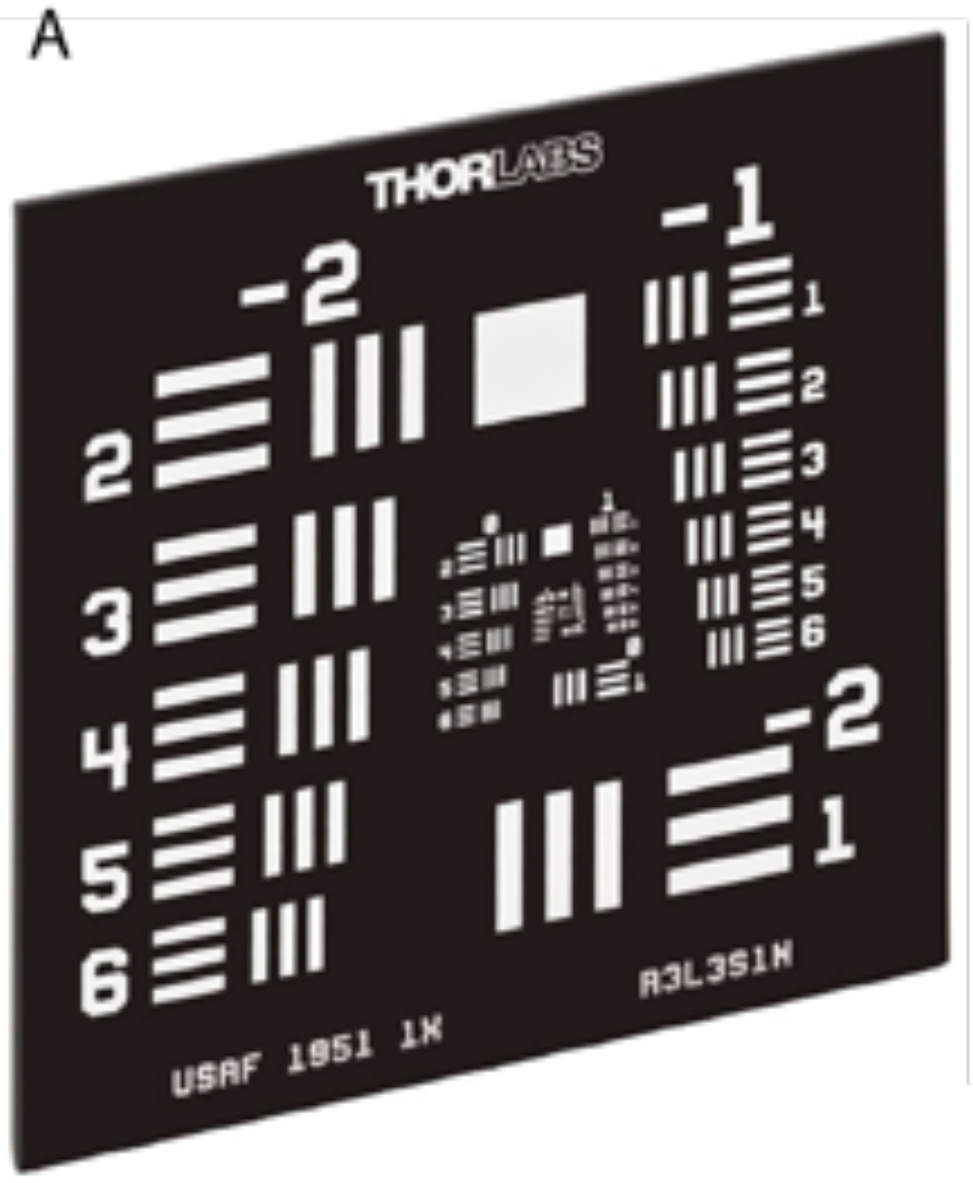

B

Resolution for USAF 1951 Test Target (Lines per Millimeter)

| Element | Group Number -2 | -1 | 0 | 1 | 2 | 3 | 4 | 5 | 6 | 7 |
|---|---|---|---|---|---|---|---|---|---|---|
| 1 | 0.25 | 0.5 | 1.00 | 2.00 | 4.00 | 8.00 | 16.00 | 32 | 64 | 128 |
| 2 | 0.28 | 0.56 | 1.12 | 2.24 | 4.49 | 8.98 | 17.95 | 36 | 72 | 144 |
| 3 | 0.32 | 0.63 | 1.26 | 2.52 | 5.04 | 10.1 | 20.16 | 40 | 81 | 161 |
| 4 | 0.35 | 0.71 | 1.41 | 2.83 | 5.66 | 11.3 | 22.62 | 45 | 91 | 181 |
| 5 | 0.4 | 0.79 | 1.59 | 3.17 | 6.35 | 12.7 | 25.39 | 51 | 102 | 203 |
| 6 | 0.45 | 0.89 | 1.78 | 3.56 | 7.13 | 14.3 | 28.50 | 57 | 114 | 228 |

*Fig. S12:* ***A.*** *Resolution Target Chart USAF 1951* ***B.*** *Resolution Characterization chart for the resolution target.*

Because these targets feature sets of three lines, they reduce the occurrence of spurious resolution and thus help prevent inaccurate resolution measurements.

**5.5. Resolution chart images for different illumination wavelengths of the 450nm – 850nm broadband MDL**

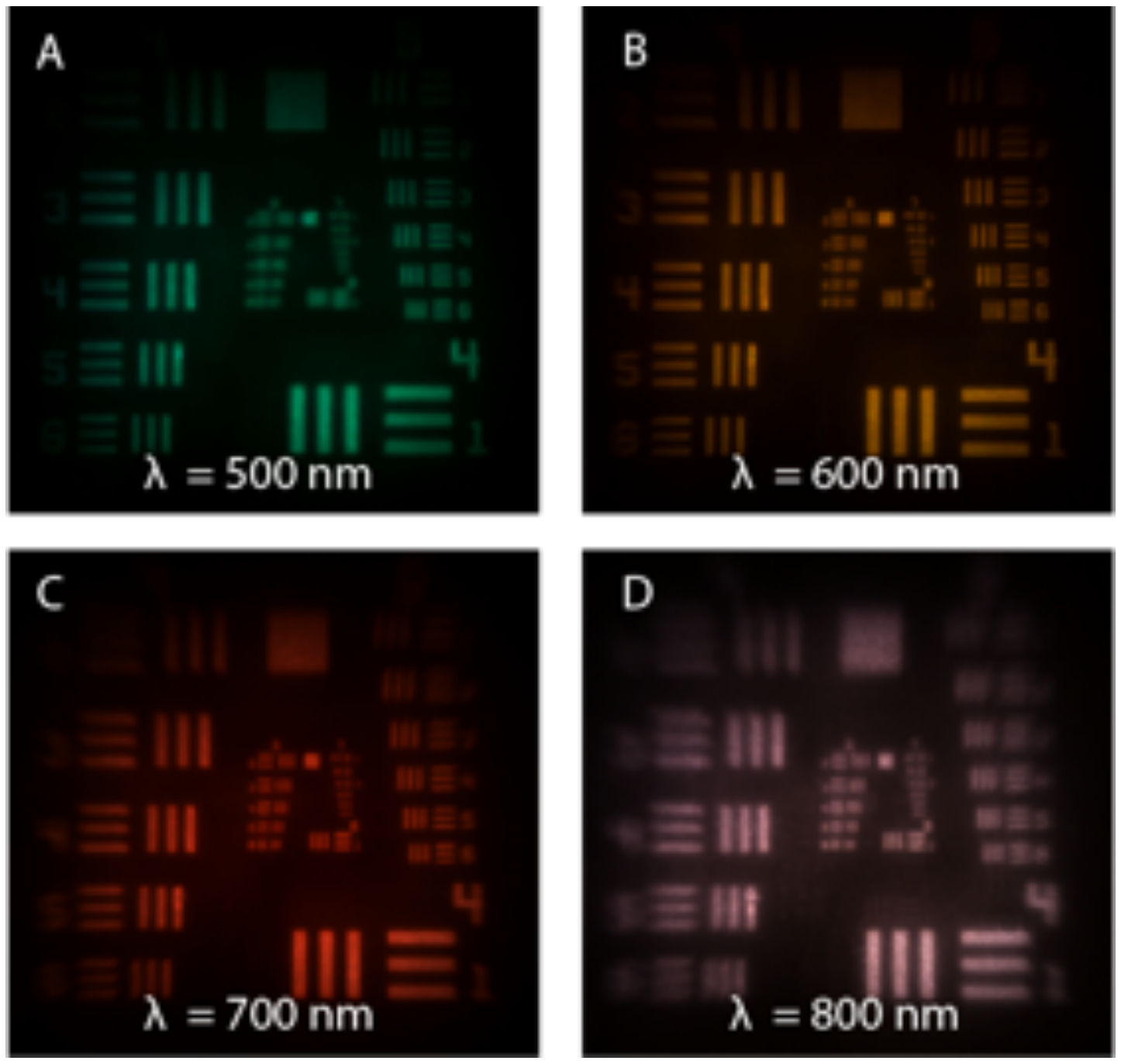

***Fig. S13:*** *Resolution chart image for the illumination wavelength* ***A.*** *500nm* ***B.*** *600nm* ***C.*** *700nm and* ***D.*** *800nm. The bandwidth at each wavelength was 100nm.*

**6. Aberrations analysis for measured wavefront of broadband (450 nm – 850 nm) MDL**

The following optical setup, shown in Figs. S14A and B was constructed to measure the aberrations of the VIS/NIR MDL using a Shack Hartmann wavefront sensor. The incident (at wavelengths 450-850 nm) beam from a SuperK EXTREME EXW-6 source [21], connected to a SuperK VARIA filter [22] was expanded and collimated and directed using a series of optical mirrors toward the measurement setup. An iris was placed in the path of the beam to limit the beam diameter. The final beam was set to have a diameter of 25.4 mm. This beam was incident on the MDL. A Shack-Hartmann wavefront sensor from Thorlabs (WFS 150-7AR) was placed 7.7 mm (MDL to outer rim of the wavefront sensor) behind the test lens [23, 24]. The distance between the outer rim to the sensor plane is 13.6 mm, making the total distance between the MDL being tested and the wavefront sensor to be 21.3 mm. The incoming collimated beam converges to a focal spot at ~ 1 mm behind the MDL and then diverges, before contacting the wavefront sensor. Hence, it is expected that the sensor should show a diverging wavefront. We selected 2.3 mm as the diameter of the pupil of the Shack Hartmann wavefront sensor (SH WFS). This ensures that neither is the beam clipped, nor oversampled. Similar experiment was performed with a stock refractive singlet lens as shown in Fig. S14A and the aberrations of the stock lens compared with those corresponding to our MDL.

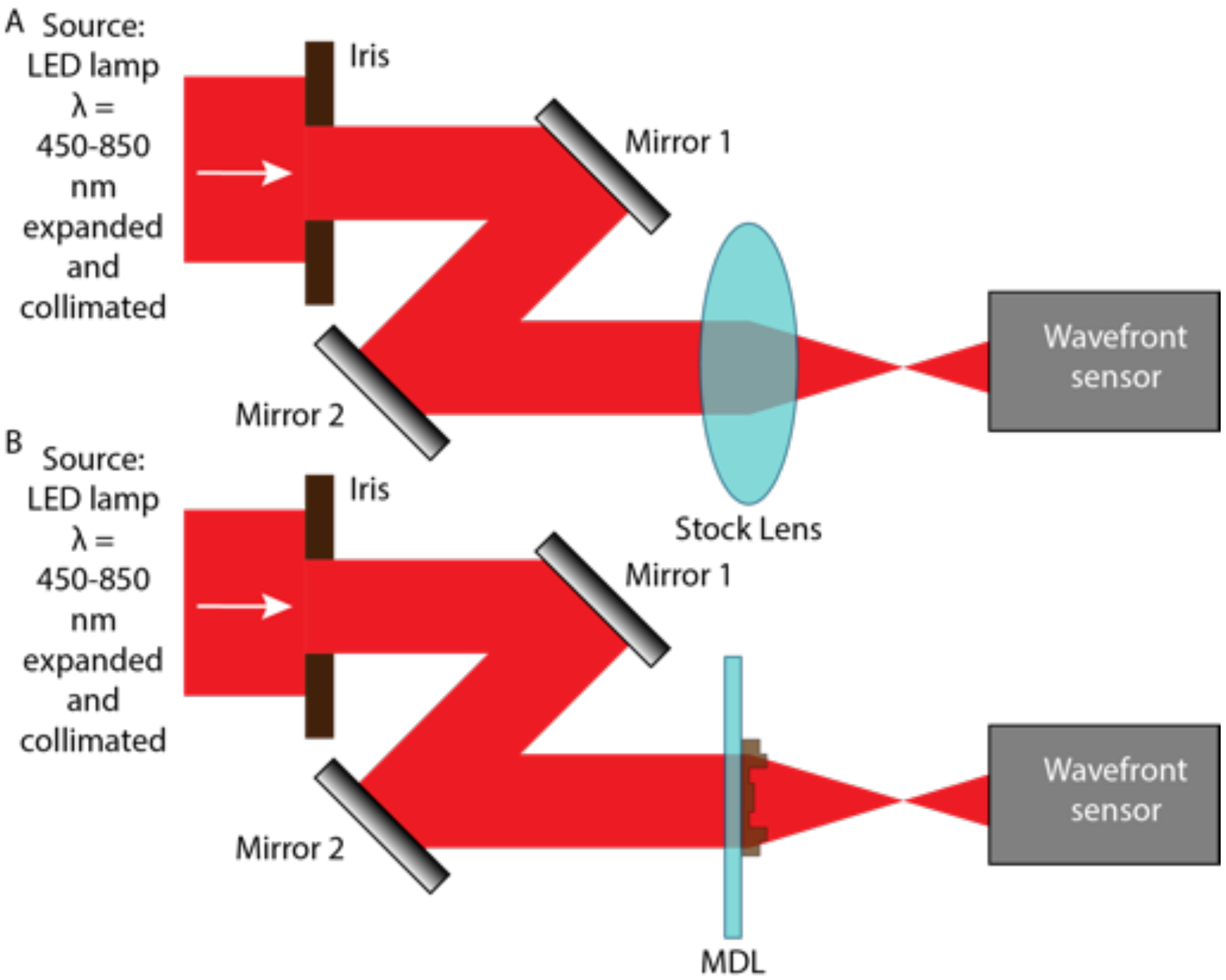

***Fig. S14****: Optical setup used to measure aberrations of a test VIS/NIR multi-level diffractive lens. The wavefront sensor is illuminated by a diverging wavefront produced after a collimated beam is brought into focus ~ 1mm behind the MDL. A series of mirrors, iris and lens are used to relay the beam from the source to the MDL. Setup for testing a (****A****) stock refractive singlet lens and (****B****)our MDL.*

First, we used a stock refractive singlet lens (plano-convex lens, Thorlabs) to test and calibrate the setup. The measurements from this are shown in Fig. S15. The alignment is confirmed to be good from the results obtained. Focal length of the lens as obtained from the SH WFS is checked with the know focal length. The known focal length of the lens was 200 mm as specified by the manufacturer. We measured the focal length to be 200.86 mm under the broadband illumination (450-850 nm) and 200.86 mm under 600 nm illumination based on the data gathered from the SH WFS. This step serves as a calibration step and the values of the stock lens serve as ground truth. Later, we compared the aberration values obtained for the MDL against this stock lens. Next, the MDL was placed in the location of the stock lens and the measurement was repeated. The measured focal lengths of the MDL were 0.988 mm under the broadband illumination (450-850 nm) and 1.06 mm under 600 nm illumination based on the data gathered from the SH WFS. The wavefronts for the stock lens and the MDL under various illumination conditions are presented in Fig. S15.

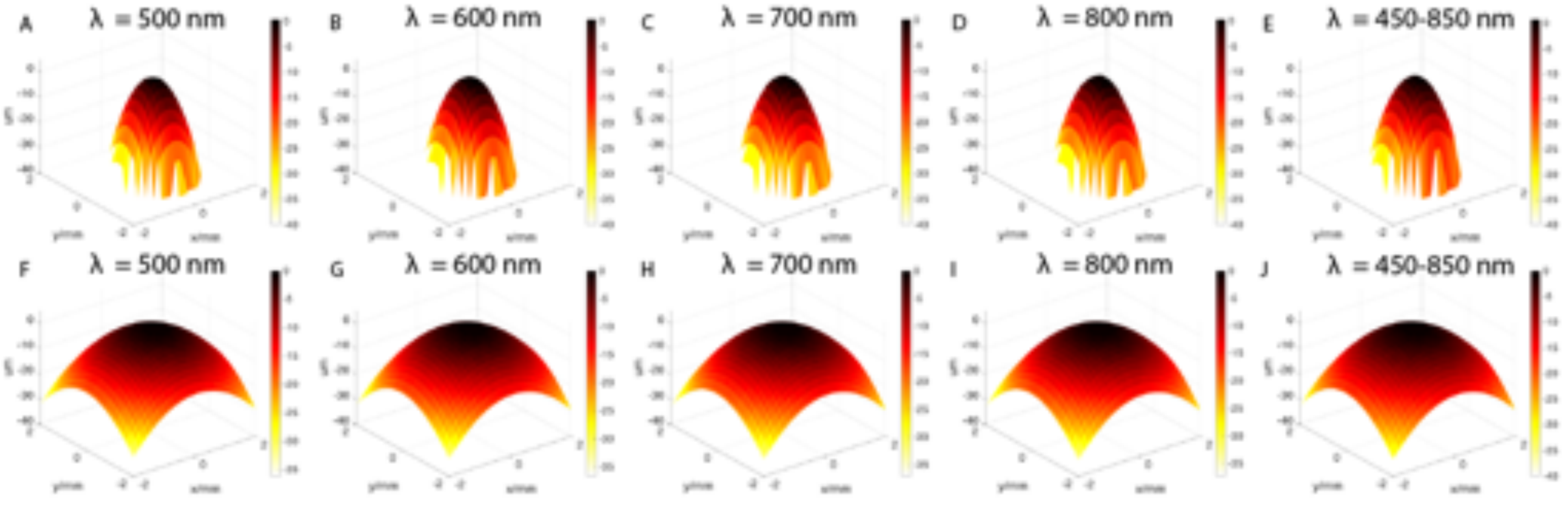

***Fig. S15**: Wavefronts for (**A-E**) for the MDL and (**F-J**) for the stock refractive singlet lens under (**A-D and F-I**) narrowband illuminations at 500, 600, 700 and 800 nm with 50 nm bandwidth and (**E and J**) broadband illumination at 450-850 nm The wavefronts of the stock lens were used as ground truth and for system calibration.*

**Table S2.** Measured aberration values for the test MDL lens and the stock refractive singlet lens

| Aberration | Test MDL Value (µm) | | | | | Stock refractive singlet lens Value (µm) | | | | |
|---|---|---|---|---|---|---|---|---|---|---|
| | λ = 500 nm | λ = 600 nm | λ = 700 nm | λ = 800 nm | λ = 450-850 nm | λ = 500 nm | λ = 600 nm | λ = 700 nm | λ = 800 nm | λ = 450-850 nm |
| Piston | 7.039 | 6.904 | 7.077 | 6.969 | 4.981 | 5.678 | 5.784 | 5.762 | 5.802 | 5.777 |
| Tip Y | 1.467 | 1.503 | 0.508 | 0.595 | 2.822 | -0.575 | -0.605 | -0.561 | -0.582 | -0.62 |
| Tilt X | 0.199 | 0.234 | 0.945 | 0.877 | 0.831 | 0.438 | 0.433 | 0.471 | 0.478 | 0.471 |
| Astigmatism ±45° | 0.013 | 0.003 | 0.02 | 0.011 | 0.008 | 0.02 | 0.022 | 0.022 | 0.021 | 0.019 |
| Defocus | -8.684 | -8.537 | -8.565 | -8.495 | -8.592 | -5.628 | -5.708 | -5.771 | -5.797 | -5.72 |
| Astigmatism 0/90° | 0.008 | 0.009 | 0.008 | 0.029 | -0.009 | 0.072 | 0.074 | 0.078 | 0.081 | 0.076 |
| Trefoil Y | 0.001 | 0.004 | -0.028 | -0.009 | 0.001 | -0.002 | -0.003 | -0.002 | -0.003 | -0.001 |
| Coma X | -0.008 | -0.022 | 0.016 | 0.007 | 0.007 | -0.007 | -0.007 | -0.009 | -0.007 | -0.008 |
| Coma Y | -0.014 | -0.009 | 0.008 | 0.006 | 0.004 | 0.008 | 0.01 | 0.011 | 0.009 | 0.008 |
| Trefoil X | 0.005 | 0.009 | -0.009 | -0.002 | -0.001 | -0.005 | -0.002 | -0.003 | -0.004 | -0.005 |
| Tetrafoil Y | 0.007 | 0.001 | 0.001 | 0.005 | 0 | -0.003 | -0.003 | -0.003 | -0.002 | -0.003 |
| Sec. Astig. Y | -0.002 | 0.003 | -0.005 | 0 | 0.002 | 0.001 | 0.002 | 0.001 | 0.001 | 0.001 |
| Spher. Aberr. 3rd O | -0.016 | -0.111 | 0.032 | -0.018 | -0.053 | 0.002 | 0.002 | 0.004 | 0.003 | 0.004 |
| Sec. Astig. X | -0.003 | -0.003 | -0.003 | -0.003 | 0.004 | 0.001 | 0 | 0 | 0 | 0 |
| Terafoil X | -0.013 | 0.013 | -0.006 | 0.027 | -0.006 | 0 | -0.001 | 0.001 | 0 | 0.001 |

## 7. Aberrations analysis for simulated wavefront

The Zernike polynomial coefficient were fitted over a circular shaped pupil. The calculation was done using the reference. The fit was using the least squares fit method.  Fringe indexing scheme was used.

**Table S3:** Aberrations coefficients

| Radial degree (n) | Azimuthal degree (m) | Fringe index (j) | Classical name |
|---|---|---|---|
| 0 | 0 | 1 | piston |
| 1 | 1 | 2 | tip |
| 1 | -1 | 3 | tilt |
| 2 | 0 | 4 | defocus |
| 2 | 2 | 5 | vertical astigmatism |
| 2 | -2 | 6 | oblique astigmatism |
| 3 | 1 | 7 | horizontal coma |
| 3 | -1 | 8 | vertical coma |
| 4 | 0 | 9 | primary spherical |
| 3 | 3 | 10 | oblique trefoil |
| 3 | -3 | 11 | vertical trefoil |
| 4 | 2 | 12 | vertical secondary astigmatism |
| 4 | -2 | 13 | oblique secondary astigmatism |
| 4 | 4 | 14 | vertical quadrafoil |
| 4 | -4 | 15 | oblique quadrafoil |

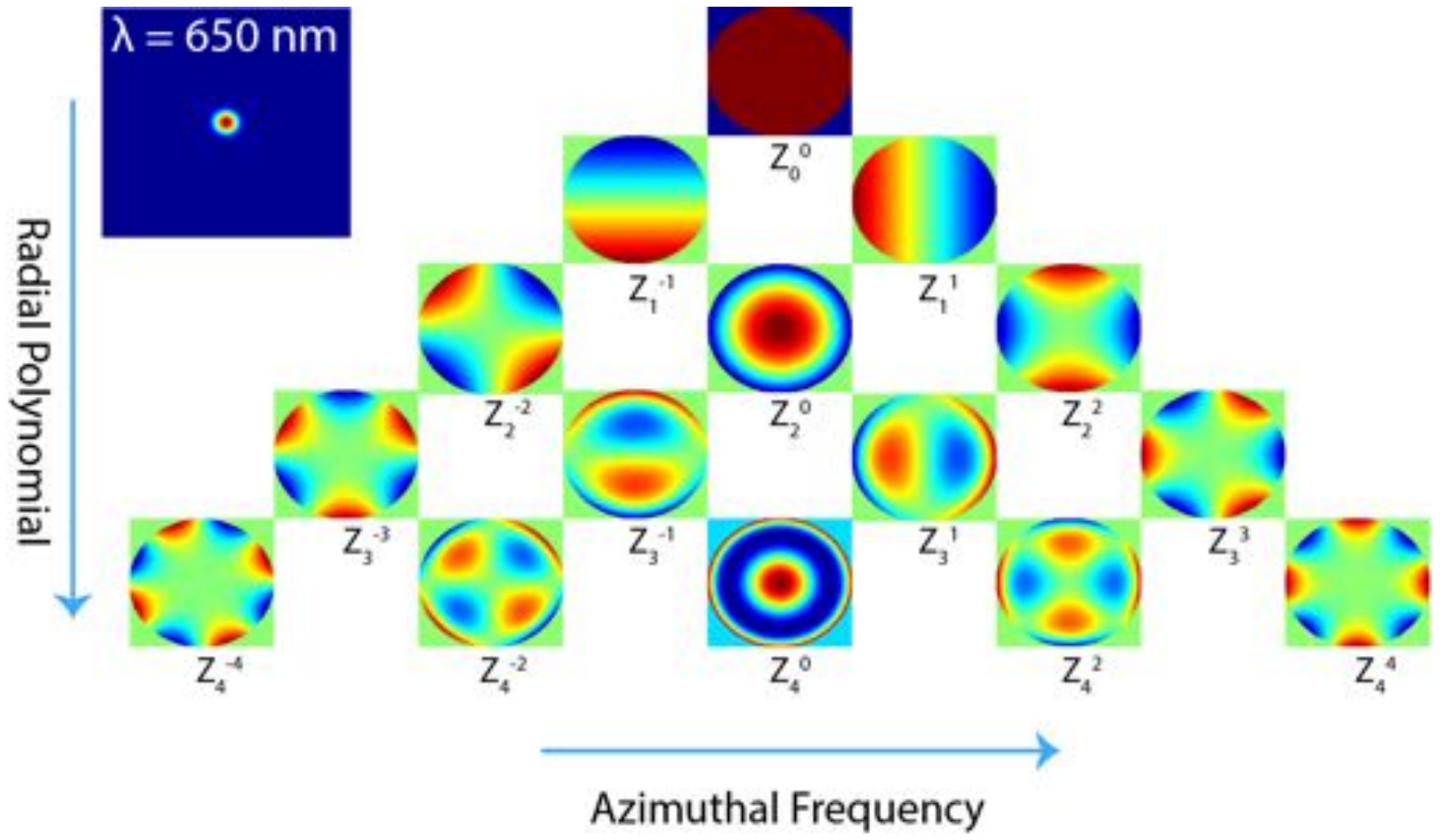

***Fig. S16:*** *Simulated wavefront analysis using Zernike coefficients for broadband MDL at λ = 650 nm*

**Table S4:** Broadband (450 nm – 850 nm) MDL with NA = 0.075, f = 1 mm

| Wave lengt | Piston | Ti p | Til t | Defocu s | Verti cal astig | Obliq ue astig | Hori zont al | Ver tica l | Prima ry | Obl iqu e | Ver tica l | Verti cal secon | Obliq ue secon | Vert ical quad | Obli que quad |
|---|---|---|---|---|---|---|---|---|---|---|---|---|---|---|---|

| h (nm) | | | | | matism | matism | coma | coma | spherical | trefoil | trefoil | dary astigmatism | dary astigmatism | rafoil | rafoil |
|---|---|---|---|---|---|---|---|---|---|---|---|---|---|---|---|
| 450 | 1.89E-04 | -2.76E-07 | -2.76E-07 | -0.0005284113 | 1.69E-11 | -1.10E-09 | 1.09E-06 | 1.09E-06 | 0.00076652743 | 6.89E-10 | -6.89E-10 | 8.37E-11 | 5.42E-09 | 4.51E-07 | -8.94E-16 |
| 475 | 0.00028775822 | -3.17E-07 | -3.17E-07 | -0.00080884405 | 9.52E-12 | -1.26E-09 | 1.23E-06 | 1.23E-06 | 0.0011895377 | 3.13E-09 | -3.12E-09 | 6.56E-11 | 6.05E-09 | 8.37E-07 | 1.79E-14 |
| 500 | 0.00044647048 | -3.63E-07 | -3.63E-07 | -0.001252948 | 8.76E-12 | -1.44E-09 | 1.45E-06 | 1.45E-06 | 0.0018368324 | 3.90E-09 | -3.90E-09 | 2.82E-11 | 7.22E-09 | 1.49E-06 | 4.66E-14 |
| 525 | 0.00043882147 | -2.93E-07 | -2.93E-07 | -0.0012238218 | 3.86E-12 | -1.17E-09 | 1.26E-06 | 1.26E-06 | 0.0017717417 | 1.26E-09 | -1.26E-09 | 1.42E-11 | 6.45E-09 | 1.87E-06 | -6.06E-14 |
| 550 | 0.00036772378 | -2.34E-07 | -2.34E-07 | -0.0010180837 | 4.78E-12 | -9.42E-10 | 1.09E-06 | 1.09E-06 | 0.0014532738 | -4.40E-10 | 4.40E-10 | 1.44E-11 | 5.68E-09 | 2.14E-06 | 3.31E-14 |
| 575 | 0.00033477173 | -2.21E-07 | -2.21E-07 | -0.00092151144 | 4.42E-12 | -8.88E-10 | 1.08E-06 | 1.08E-06 | 0.0013021866 | 1.42E-09 | -1.42E-09 | 3.61E-12 | 5.68E-09 | 2.63E-06 | 1.99E-14 |
| 600 | 0.00032281689 | -2.19E-07 | -2.19E-07 | -0.00088604301 | 8.08E-12 | -8.58E-10 | 1.12E-06 | 1.12E-06 | 0.0012478773 | 1.42E-08 | -1.42E-08 | -1.10E-11 | 5.93E-09 | 3.22E-06 | 1.17E-14 |
| 625 | 0.00030532011 | -2.10E-07 | -2.10E-07 | -0.00083823217 | 7.81E-13 | -7.92E-10 | 1.08E-06 | 1.08E-06 | 0.0011845863 | 2.15E-08 | -2.15E-08 | -8.23E-12 | 5.76E-09 | 3.59E-06 | 8.18E-14 |
| 650 | 0.00027390639 | -1.98E | -1.98E | -0.00075407029 | -6.17E-12 | -8.23E-10 | 9.16E-07 | 9.16E-07 | 0.0010747632 | -4.60E-09 | 4.60E-09 | 1.32E-11 | 4.76E-09 | 3.61E-06 | 1.97E-14 |
| | | -07 | -07 | | | | | | | | | | | | |
| 675 | 0.00023984583 | -1.63E-07 | -1.63E-07 | -0.00066312635 | 2.78E-12 | -6.26E-10 | 7.46E-07 | 7.47E-07 | 0.00095563574 | 1.98E-09 | -1.98E-09 | 1.87E-12 | 3.92E-09 | 3.42E-06 | 2.67E-14 |
| 700 | 0.00021409358 | -1.34E-07 | -1.34E-07 | -0.00059477816 | 4.90E-12 | -5.34E-10 | 5.97E-07 | 5.97E-07 | 0.0008669591 | 3.83E-09 | -3.83E-09 | -2.94E-11 | 3.06E-09 | 3.22E-06 | 1.12E-14 |
| 725 | 0.00019938 | -1.12E-07 | -1.12E-07 | -0.00055640505 | 2.45E-12 | -4.39E-10 | 4.90E-07 | 4.90E-07 | 0.00081926677 | 6.39E-09 | -6.39E-09 | -4.37E-12 | 2.50E-09 | 3.12E-06 | -1.01E-14 |
| 750 | 0.00019441386 | -1.02E-07 | -1.02E-07 | -0.00054452999 | 4.93E-12 | -3.92E-10 | 4.12E-07 | 4.12E-07 | 0.00080814079 | 7.56E-10 | -7.55E-10 | -5.76E-12 | 2.10E-09 | 3.15E-06 | -8.09E-14 |
| 775 | 0.00019743654 | -9.20E-08 | -9.20E-08 | -0.00055435608 | 5.41E-12 | -3.74E-10 | 3.75E-07 | 3.75E-07 | 0.00082709669 | 6.37E-09 | -6.37E-09 | -7.95E-12 | 1.86E-09 | 3.33E-06 | -1.59E-14 |
| 800 | 0.00020730276 | -9.00E-08 | -9.00E-08 | -0.0005827109 | -7.27E-14 | -3.41E-10 | 3.59E-07 | 3.59E-07 | 0.00087170815 | 5.87E-09 | -5.87E-09 | -2.29E-12 | 1.82E-09 | 3.65E-06 | 6.94E-15 |
| 825 | 0.00022363955 | -9.39E-08 | -9.38E-08 | -0.00062846224 | 2.60E-12 | -3.51E-10 | 3.66E-07 | 3.66E-07 | 0.00094021123 | 3.32E-09 | -3.32E-09 | -1.11E-11 | 1.88E-09 | 4.16E-06 | -2.08E-14 |
| 850 | 0.00024661465 | -1.04E-07 | -1.04E-07 | -0.0006918846 | -4.00E-12 | -4.63E-10 | 4.00E-07 | 4.00E-07 | 0.0010326691 | -1.44E-09 | 1.44E-09 | 1.16E-11 | 1.96E-09 | 4.86E-06 | 3.61E-14 |

**Table S5:** Broadband (0.5μm –15μm) MDL with NA = 0.0797, f = 5 mm

| **Waveleng** | **Piston** | **Tip** | **Tilt** | **Defocus** | **Vertical astig** | **Oblique astig** | **Horizontal** | **Vertical** | **Primary** | **Oblique** | **Vertical** | **Vertical secon** | **Oblique secon** | **Vertical** | **Oblique qua** |
|---|---|---|---|---|---|---|---|---|---|---|---|---|---|---|---|

| **th (um)** | | | | | **matism** | **matism** | **coma** | **coma** | **spherical** | **trefoil** | **trefoil** | **dary astigmatism** | **dary astigmatism** | **quadrafoil** | **drafoil** |
|---|---|---|---|---|---|---|---|---|---|---|---|---|---|---|---|
| 0.2 | 0.043129627 | 5.33E-05 | 5.33E-05 | -0.00051947299 | 1.54E-08 | 1.24E-07 | -2.33E-05 | -2.33E-05 | 0.0015698343 | -1.30E-04 | 1.30E-04 | 3.70E-09 | 2.54E-06 | 0.004240674 | -1.75E-10 |
| 0.25 | 0.050028447 | -1.71E-05 | -1.70E-05 | -0.0052477336 | 8.97E-09 | 2.96E-06 | -9.29E-05 | -9.29E-05 | 0.011984387 | -7.17E-05 | 7.17E-05 | -8.55E-09 | 4.65E-06 | 1.84E-02 | -9.36E-10 |
| 0.3 | 0.022603258 | -1.74E-05 | -1.74E-05 | 0.0084305527 | 1.74E-09 | -1.12E-06 | -1.19E-05 | -1.19E-05 | 0.0028690433 | -2.83E-05 | 2.83E-05 | -7.08E-09 | -2.39E-06 | 0.005442712 | 3.40E-10 |
| 0.35 | 0.024987604 | -1.90E-05 | -1.90E-05 | -0.014578459 | -2.58E-10 | 4.33E-07 | 4.90E-06 | 4.89E-06 | 5.90E-05 | -3.90E-05 | 3.90E-05 | -4.99E-09 | 6.50E-07 | -6.77E-03 | -1.47E-10 |
| 0.4 | 0.012688935 | -2.59E-05 | -2.59E-05 | -0.00094959908 | 4.62E-09 | -4.52E-07 | 9.49E-06 | 9.48E-06 | -0.00089047517 | -6.20E-06 | 6.20E-06 | -2.64E-10 | -6.35E-07 | -1.20E-03 | -3.81E-10 |
| 0.45 | 0.0058505274 | -1.10E-05 | -1.10E-05 | 0.00042408373 | -1.55E-09 | -7.16E-06 | -2.46E-05 | -2.46E-05 | 0.00069258973 | 2.07E-05 | -2.07E-05 | 2.22E-11 | -1.21E-05 | 3.84E-04 | 5.60E-11 |
| 0.5 | 0.004202839 | -7.22E-06 | -7.22E-06 | 0.0015095581 | -6.70E-10 | -6.09E-07 | -1.58E-05 | -1.58E-05 | 0.001563885 | -3.59E-06 | 3.59E-06 | 2.29E-10 | -1.04E-06 | 3.55E-03 | -4.48E-11 |
| 0.55 | 0.0027676001 | 2.54E-06 | 2.54E-06 | 0.0012221405 | -1.37E-09 | -1.92E-07 | 2.87E-06 | 2.87E-06 | 0.00094057049 | 1.55E-05 | -1.55E-05 | 2.47E-10 | -3.60E-07 | 3.38E-03 | 3.27E-11 |
| 0.6 | 0.0037813571 | 6.31E-06 | 6.31E-06 | 0.0019563679 | 6.30E-10 | -5.14E-07 | 5.24E-06 | 5.24E-06 | 0.0025530409 | 2.00E-05 | -2.00E-05 | -1.23E-09 | -7.95E-07 | 5.67E-03 | -1.43E-11 |
| 0.7 | 0.0014831945 | 8.77E-07 | 8.75E-07 | -0.00043950809 | 6.74E-10 | -6.27E-08 | 4.35E-06 | 4.36E-06 | 0.0015096597 | -9.50E-07 | 9.48E-07 | 6.39E-11 | -9.06E-08 | 3.16E-03 | -8.51E-12 |
| 0.8 | 0.00062856846 | -2.52E-07 | -2.51E-07 | -0.00095389993 | 1.39E-10 | 1.04E-08 | 3.84E-06 | 3.84E-06 | 0.0011569167 | -4.94E-07 | 4.94E-07 | -1.03E-10 | 4.28E-08 | 6.17E-04 | 1.12E-12 |
| 0.9 | 0.00032157678 | -1.14E-08 | -1.12E-08 | -0.00078097585 | 5.17E-12 | 1.28E-09 | 3.43E-07 | 3.42E-07 | 0.00089705153 | 1.23E-07 | -1.23E-07 | -4.88E-11 | -2.72E-09 | 1.35E-05 | -2.17E-13 |
| 1 | 0.00022917353 | -2.29E-07 | -2.29E-07 | -0.00055989635 | 5.84E-12 | -8.05E-10 | 9.06E-07 | 9.05E-07 | 0.00060925644 | -3.23E-09 | 3.24E-09 | 2.94E-11 | 4.07E-09 | -3.61E-06 | -9.59E-14 |
| 1.1 | 0.00019577034 | -2.33E-07 | -2.33E-07 | -0.00047079852 | 9.81E-12 | -8.00E-10 | 7.24E-07 | 7.24E-07 | 0.00049949845 | -8.07E-08 | 8.07E-08 | -4.62E-11 | 2.35E-09 | -2.95E-06 | 9.77E-14 |
| 1.5 | 0.00043804504 | -4.19E-07 | -4.19E-07 | -0.0010468392 | -1.06E-11 | -1.68E-09 | 1.39E-06 | 1.39E-06 | 0.0011448755 | -2.00E-07 | 2.00E-07 | 9.53E-11 | 6.19E-09 | -4.14E-06 | -1.78E-13 |
| 2 | 0.0003521108 | 2.24E-07 | 2.24E-07 | -0.00085951824 | 6.14E-12 | 2.11E-09 | -3.33E-07 | -3.33E-07 | 0.0010705724 | 1.74E-07 | -1.74E-07 | -4.18E-11 | -7.64E-09 | -1.23E-06 | -9.90E-14 |
| 2.5 | 0.00041702986 | -8.33E-08 | -8.34E-08 | -0.00094779214 | 2.09E-11 | 1.90E-10 | 1.60E-06 | 1.60E-06 | 0.0011126674 | 2.69E-07 | -2.69E-07 | -4.93E-11 | 5.94E-09 | 7.76E-06 | -2.25E-13 |
| 3 | 0.0010581879 | -4.01E-06 | -4.01E-06 | -0.0023170155 | 1.20E-10 | -2.01E-08 | 1.40E-05 | 1.40E-05 | 0.0024887063 | -4.15E-07 | 4.15E-07 | -8.33E-11 | 4.52E-08 | -3.97E-06 | -6.56E-13 |
| 3.5 | 0.00081607845 | -1.63E-06 | -1.63E-06 | -0.0017128872 | -3.22E-11 | 2.44E-09 | 5.53E-06 | 5.53E-06 | 0.0018225836 | -3.37E-07 | 3.37E-07 | -1.88E-10 | 3.25E-08 | 1.84E-05 | 1.19E-12 |
| 4 | 0.0016434233 | -1.64E-06 | -1.64E-06 | -0.0033410373 | -1.38E-11 | -2.77E-08 | 4.84E-06 | 4.84E-06 | 0.0035245351 | -1.55E-06 | 1.55E-06 | -1.45E-10 | -3.21E-08 | -1.04E-05 | -1.24E-12 |
| 4.5 | 0.0017364344 | -8.67E-08 | -8.62E-08 | -0.0036120436 | -1.28E-11 | 1.00E-08 | -4.49E-07 | -4.48E-07 | 0.0040642773 | -1.32E-06 | 1.32E-06 | 4.64E-10 | -3.08E-08 | -6.82E-05 | 4.52E-12 |
| 5 | 0.0019421463 | 1.37E-06 | 1.37E-06 | -0.0041691996 | -6.33E-11 | 1.15E-08 | -6.40E-06 | -6.40E-06 | 0.005136448 | -1.03E-06 | 1.03E-06 | -4.05E-11 | -1.01E-07 | -7.44E-05 | 1.99E-13 |
| 6 | 0.0027851001 | 4.45E-06 | 4.45E-06 | -0.0065798601 | -6.52E-11 | -3.62E-09 | -1.32E-05 | -1.32E-05 | 0.0089313881 | 4.85E-06 | -4.85E-06 | 2.48E-10 | -2.60E-07 | -1.03E-04 | -1.68E-12 |
| 7 | 0.0053723757 | 3.80E-06 | 3.80E-06 | -0.012821789 | -1.19E-10 | 6.11E-08 | -7.31E-06 | -7.32E-06 | 0.016866619 | 7.84E-06 | -7.84E-06 | -5.67E-10 | -3.72E-07 | -0.00023451603 | -4.49E-12 |
| 8 | 0.0048783063 | -2.21E-06 | -2.21E-06 | -0.011565058 | -1.46E-10 | 4.53E-08 | 2.17E-06 | 2.17E-06 | 0.014858129 | 3.56E-06 | -3.56E-06 | -1.52E-10 | -3.14E-07 | -1.15E-04 | -1.22E-12 |
| 9 | 0.002642727 | 2.28E-06 | 2.28E-06 | -0.0064547104 | -1.59E-10 | 3.15E-08 | -1.16E-05 | -1.16E-05 | 0.0088611003 | 9.66E-07 | -9.66E-07 | -1.99E-10 | -1.73E-07 | -7.49E-05 | 1.48E-12 |
| 10 | 0.0024063708 | 3.52E-06 | 3.52E-06 | -0.0058758785 | -5.27E-11 | 4.22E-08 | -1.29E-05 | -1.29E-05 | 0.0081241317 | 2.17E-06 | -2.17E-06 | 1.09E-11 | -1.52E-07 | -7.83E-05 | -2.12E-12 |
| 11 | 0.0028647333 | 4.36E-06 | 4.36E-06 | -0.0069442904 | 1.23E-10 | 4.82E-08 | -1.77E-05 | -1.77E-05 | 0.0094127292 | 1.64E-06 | -1.64E-06 | -4.81E-12 | -2.00E-07 | 1.49E-06 | -4.98E-13 |
| 12 | 0.003659429 | 4.28E-06 | 4.28E-06 | -0.0087825619 | -2.59E-10 | 4.90E-08 | -2.36E-05 | -2.36E-05 | 0.011808945 | -8.02E-07 | 8.02E-07 | -6.29E-10 | -2.70E-07 | 5.30E-05 | -1.27E-13 |
| 13 | 0.0049489462 | 3.60E-06 | 3.60E-06 | -0.011765971 | -1.89E-10 | 2.70E-08 | -3.12E-05 | -3.12E-05 | 0.015697466 | -2.11E-06 | 2.11E-06 | 9.73E-11 | -4.36E-07 | 2.58E-05 | 4.81E-13 |
| 14 | 0.0074428599 | 9.45E-06 | 9.44E-06 | -0.017742429 | 3.98E-10 | 6.26E-08 | -4.96E-05 | -4.96E-05 | 0.023715939 | 2.85E-06 | -2.85E-06 | -5.21E-10 | -6.56E-07 | 3.73E-05 | -1.46E-11 |
| 15 | 0.011615873 | 2.21E-05 | 2.21E-05 | -0.027628809 | -8.25E-11 | 2.92E-07 | -6.61E-05 | -6.61E-05 | 0.0362649 | 1.53E-05 | -1.53E-05 | 1.08E-09 | -7.18E-07 | 1.94E-04 | 4.01E-12 |

**Table S6:** Broadband (2μm– 150μm) MDL with NA = 0.0998, f = 10 mm

| **Wavelength (um)** | **Piston** | **Tip** | **Tilt** | **Defocus** | **Vertical astigmatism** | **Oblique astigmatism** | **Horizontal coma** | **Vertical coma** | **Primary spherical** | **Oblique trefoil** | **Vertical trefoil** | **Vertical secondary astigmatism** | **Oblique secondary astigmatism** | **Vertical quadrafoil** | **Oblique quadrafoil** |
|---|---|---|---|---|---|---|---|---|---|---|---|---|---|---|---|
| 2.00E-06 | 3.60E-05 | -4.11E-09 | -4.11E-09 | -8.76E-05 | 1.64E-11 | -4.22E-12 | 1.98E-08 | 1.98E-08 | 0.00010411395 | -1.27E-09 | 1.27E-09 | 4.57E-11 | 1.53E-11 | -4.50E-07 | 7.03E-15 |
| 2.50E-06 | 3.07E-05 | -1.06E-08 | -1.07E-08 | -6.97E-05 | 1.82E-11 | -2.50E-11 | 2.12E-08 | 2.12E-08 | 8.54E-05 | -1.18E-08 | 1.18E-08 | -2.54E-12 | 7.99E-12 | 8.34E-08 | 6.31E-15 |
| 3.00E-06 | 3.34E-05 | -2.74E-08 | -2.75E-08 | -7.13E-05 | -3.19E-12 | -1.45E-11 | 2.04E-08 | 2.04E-08 | 9.35E-05 | -2.90E-08 | 2.90E-08 | -7.25E-12 | 6.47E-11 | 1.43E-07 | 9.96E-15 |
| 3.50E-06 | 5.49E-05 | 5.04E-09 | 5.05E-09 | -0.00011819089 | 1.97E-11 | 1.70E-11 | 2.33E-08 | 2.33E-08 | 0.00014952042 | 1.55E-08 | -1.56E-08 | -4.56E-11 | -6.56E-11 | 3.00E-07 | 3.15E-14 |
| 4.00E-06 | 7.48E-05 | -3.10E-08 | -3.10E-08 | -0.00015031143 | -8.13E-12 | 4.30E-11 | -1.70E-08 | -1.69E-08 | 0.00019822488 | -2.65E-09 | 2.65E-09 | 4.40E-12 | -1.17E-10 | -1.60E-07 | -1.48E-14 |
| 4.50E-06 | 7.71E-05 | -9.58E-09 | -9.55E-09 | -0.00016342379 | -1.19E-11 | -3.13E-10 | -6.19E-09 | -6.18E-09 | 0.00019971747 | 1.71E-09 | -1.72E-09 | 5.03E-12 | -7.16E-10 | -6.99E-07 | 1.32E-14 |
| 5.00E-06 | 0.00011358057 | -4.54E-08 | -4.54E-08 | -0.00025202581 | 2.64E-11 | 1.03E-09 | 1.46E-07 | 1.46E-07 | 0.00031575485 | 5.98E-09 | -5.96E-09 | -5.67E-11 | 1.93E-09 | -5.18E-07 | -7.54E-14 |
| 6.00E-06 | 0.00020706403 | -9.05E-08 | -9.08E-08 | -0.0004453591 | 9.53E-11 | -1.03E-10 | 3.44E-07 | 3.44E-07 | 0.00055458734 | -2.81E-08 | 2.82E-08 | -3.90E-11 | 3.55E-10 | -2.05E-06 | -2.32E-13 |
| 7.00E-06 | 0.00026649286 | -1.36E-07 | -1.37E-07 | -0.00060508552 | 1.10E-10 | 7.25E-10 | 2.48E-07 | 2.48E-07 | 0.00073988643 | -7.19E-08 | 7.21E-08 | -1.52E-11 | 1.37E-09 | -2.38E-06 | 2.71E-14 |
| 8.00E-06 | 0.00024383781 | -1.32E-07 | -1.33E-07 | -0.00052641181 | 3.57E-11 | 1.36E-09 | 3.75E-07 | 3.76E-07 | 0.00066956913 | 7.41E-08 | -7.43E-08 | 3.98E-11 | 2.45E-09 | -6.79E-06 | 2.63E-13 |
| 9.00E-06 | 0.00040594983 | -2.31E-07 | -2.31E-07 | -0.00091975869 | -4.45E-12 | -1.95E-10 | 5.83E-07 | 5.84E-07 | 0.0011838074 | -4.37E-08 | 4.36E-08 | -7.83E-11 | 3.94E-10 | -1.23E-05 | 1.43E-13 |
| 1.00E-05 | 0.00044099125 | -2.22E-07 | -2.23E-07 | -0.00099092524 | 1.02E-10 | -2.21E-09 | 5.43E-07 | 5.43E-07 | 0.0012631908 | 8.62E-08 | -8.64E-08 | -1.72E-10 | -3.71E-09 | -1.47E-05 | 5.02E-14 |
| 1.10E-05 | 0.00053592486 | -2.34E-07 | -2.35E-07 | -0.001190097 | -1.78E-11 | -3.59E-09 | 7.83E-07 | 7.84E-07 | 0.0016301807 | 2.26E-07 | -2.27E-07 | 3.26E-11 | -6.32E-09 | -5.20E-05 | 1.12E-13 |
| 1.20E-05 | 0.0004964312 | -2.50E-07 | -2.50E-07 | -0.0011961684 | -4.37E-11 | -6.92E-10 | 6.19E-07 | 6.19E-07 | 0.001639573 | -3.98E-09 | 4.12E-09 | -8.78E-11 | -9.05E-10 | -3.21E-05 | 4.01E-13 |
| 1.30E-05 | 0.00050470646 | -3.36E-07 | -3.36E-07 | -0.001251628 | -1.00E-11 | -1.03E-09 | 5.18E-07 | 5.19E-07 | 0.0016592756 | -1.22E-07 | 1.23E-07 | -1.53E-10 | -9.67E-10 | -1.26E-05 | -8.37E-14 |
| 1.40E-05 | 0.00070995261 | -4.16E | -4.17E | -0.0017 | 8.88E-11 | 1.05E-09 | 9.84E-07 | 9.84E-07 | 0.0022228849 | 1.27E-09 | -1.3 | -7.87E-10 | 2.95E-09 | -5.29E-06 | 1.44E-13 |
| | | -07 | -07 | 417667 | | | | | | | 4E-09 | | | | |
| 1.50E-05 | 0.0012607493 | -8.56E-07 | -8.57E-07 | -0.0029816185 | 2.64E-10 | -9.82E-10 | 2.26E-06 | 2.26E-06 | 0.0037503445 | -1.48E-07 | 1.48E-07 | 1.43E-11 | 1.30E-09 | -3.44E-05 | 4.14E-13 |
| 2.00E-05 | 0.0012002293 | -7.69E-07 | -7.70E-07 | -0.0029419332 | 1.85E-10 | 4.87E-10 | 1.60E-06 | 1.60E-06 | 0.0038537516 | -1.09E-07 | 1.09E-07 | -1.19E-11 | 4.00E-10 | -1.31E-06 | 1.83E-12 |
| 2.50E-05 | 0.0023448709 | -1.66E-06 | -1.66E-06 | -0.0058032391 | 2.79E-10 | -2.42E-09 | 3.47E-06 | 3.47E-06 | 0.0073477505 | -4.25E-07 | 4.26E-07 | -8.08E-10 | 1.44E-09 | 9.57E-05 | -3.20E-13 |
| 3.00E-05 | 0.0034424728 | -2.61E-06 | -2.61E-06 | -0.008157203 | 7.34E-10 | -1.21E-09 | 7.90E-06 | 7.89E-06 | 0.0099418815 | -1.56E-10 | -2.85E-10 | -3.71E-10 | 9.16E-09 | 0.00013789303 | 4.34E-12 |
| 3.50E-05 | 0.0036003364 | -4.56E-06 | -4.57E-06 | -0.0081717176 | -1.15E-10 | -3.97E-09 | 7.40E-06 | 7.39E-06 | 0.0099516697 | -2.26E-06 | 2.26E-06 | 1.24E-09 | 3.59E-09 | 7.66E-05 | 1.03E-13 |
| 4.00E-05 | 0.0043393276 | -4.72E-06 | -4.72E-06 | -0.0096239634 | 2.37E-10 | -5.27E-09 | 7.49E-06 | 7.49E-06 | 0.012062714 | -1.99E-06 | 2.00E-06 | 3.32E-09 | -1.16E-08 | 2.75E-06 | -7.50E-12 |
| 4.50E-05 | 0.0064581926 | -5.33E-06 | -5.33E-06 | -0.014184326 | -1.49E-10 | 1.29E-09 | 3.83E-06 | 3.83E-06 | 0.018753756 | -2.07E-06 | 2.07E-06 | 7.03E-10 | -3.40E-08 | -0.00018034986 | 1.17E-11 |
| 5.00E-05 | 0.011248727 | -2.95E-06 | -2.95E-06 | -0.025116773 | 9.12E-10 | 2.91E-08 | -9.82E-07 | -9.78E-07 | 0.03454797 | 2.32E-06 | -2.32E-06 | -1.54E-09 | -5.13E-08 | -0.00044698146 | 4.81E-11 |
| 6.00E-05 | 0.027711842 | 1.26E-05 | 1.26E-05 | -0.063778833 | 8.90E-09 | 4.77E-08 | -3.29E-05 | -3.29E-05 | 0.089453511 | 1.72E-05 | -1.73E-05 | -1.29E-11 | -2.36E-07 | -0.00086950965 | 1.33E-11 |
| 7.00E-05 | 0.03307379 | 1.84E-05 | 1.85E-05 | -0.077880569 | 7.27E-09 | 2.09E-08 | -6.62E-05 | -6.62E-05 | 0.10969856 | 2.13E-05 | -2.14E-05 | -1.51E-08 | -3.93E-07 | 0.00021952733 | 8.12E-11 |
| 8.00E-05 | 0.038954012 | 2.73E-05 | 2.74E-05 | -0.094490349 | 4.74E-09 | 8.51E-08 | -7.45E-05 | -7.45E-05 | 0.13149491 | 3.59E-05 | -3.59E-05 | -2.33E-08 | -3.53E-07 | 0.0023152409 | -9.47E-11 |
| 9.00E-05 | 0.030772073 | 1.86E-05 | 1.86E-05 | -0.074431442 | 1.60E-08 | 6.71E-08 | -6.26E-05 | -6.26E-05 | 0.10226212 | 1.71E-05 | -1.72E-05 | 2.95E-08 | -2.35E-07 | 0.0035452121 | 4.53E-11 |
| 0.0001 | 0.027050912 | 1.52E-05 | 1.52E-05 | -0.066132225 | 1.69E-08 | 5.89E-08 | -5.59E-05 | -5.59E-05 | 0.090980023 | 1.10E-05 | -1.10E-05 | -1.15E-08 | -1.81E-07 | 0.0044395276 | 4.03E-12 |
| 0.00011 | 0.026356924 | 1.73E-05 | 1.73E-05 | -0.063480824 | 4.42E-08 | 6.44E-08 | -4.66E-05 | -4.67E-05 | 0.085929327 | 1.38E-05 | -1.38E-05 | 8.97E-09 | -1.39E-07 | 0.0053840363 | -7.69E-11 |
| 0.00012 | 0.026124706 | 1.57E-05 | 1.57E-05 | -0.063553244 | 1.58E-08 | 4.76E-08 | -3.78E-05 | -3.77E-05 | 0.086945102 | 1.04E-05 | -1.05E-05 | -1.39E-07 | -1.10E-07 | 0.0058447546 | 6.57E-11 |
| 0.00013 | 0.026890904 | 1.55E-05 | 1.55E-05 | -0.065207154 | 3.11E-08 | 4.71E-08 | -4.02E-05 | -4.02E-05 | 0.089368679 | 6.37E-06 | -6.38E-06 | -6.11E-08 | -8.42E-08 | 0.0073220516 | -1.24E-10 |
| 0.00014 | 0.028860269 | 1.56E-05 | 1.56E-05 | -0.068073899 | 1.13E-08 | 4.29E-08 | -4.77E-05 | -4.77E-05 | 0.091637485 | 4.10E-06 | -4.09E-06 | 2.37E-09 | -9.24E-08 | 0.0090332562 | -4.24E-11 |
| 0.00015 | 0.031044755 | 1.40E-05 | 1.40E-05 | -0.071999855 | 4.75E-08 | 3.99E-08 | -5.19E-05 | -5.19E-05 | 0.095432326 | 2.89E-06 | -2.90E-06 | 4.99E-08 | -9.25E-08 | 0.010147722 | 3.08E-11 |

**8. Distortion analysis for measured wavefront of broadband (450 nm – 850 nm) MDL**

The following optical setup, shown in Fig. S14 was constructed to measure the geometric distortion, vignetting and signal-to-noise ratio (SNR) of the VIS/NIR MDL. A screen with different patterns used for the different experiments was placed 115 mm away from the MDL, which in turn was placed in front of a sensor. Two white LEDS illuminate the screen from far away, so that the screen may be uniformly illuminated [25].

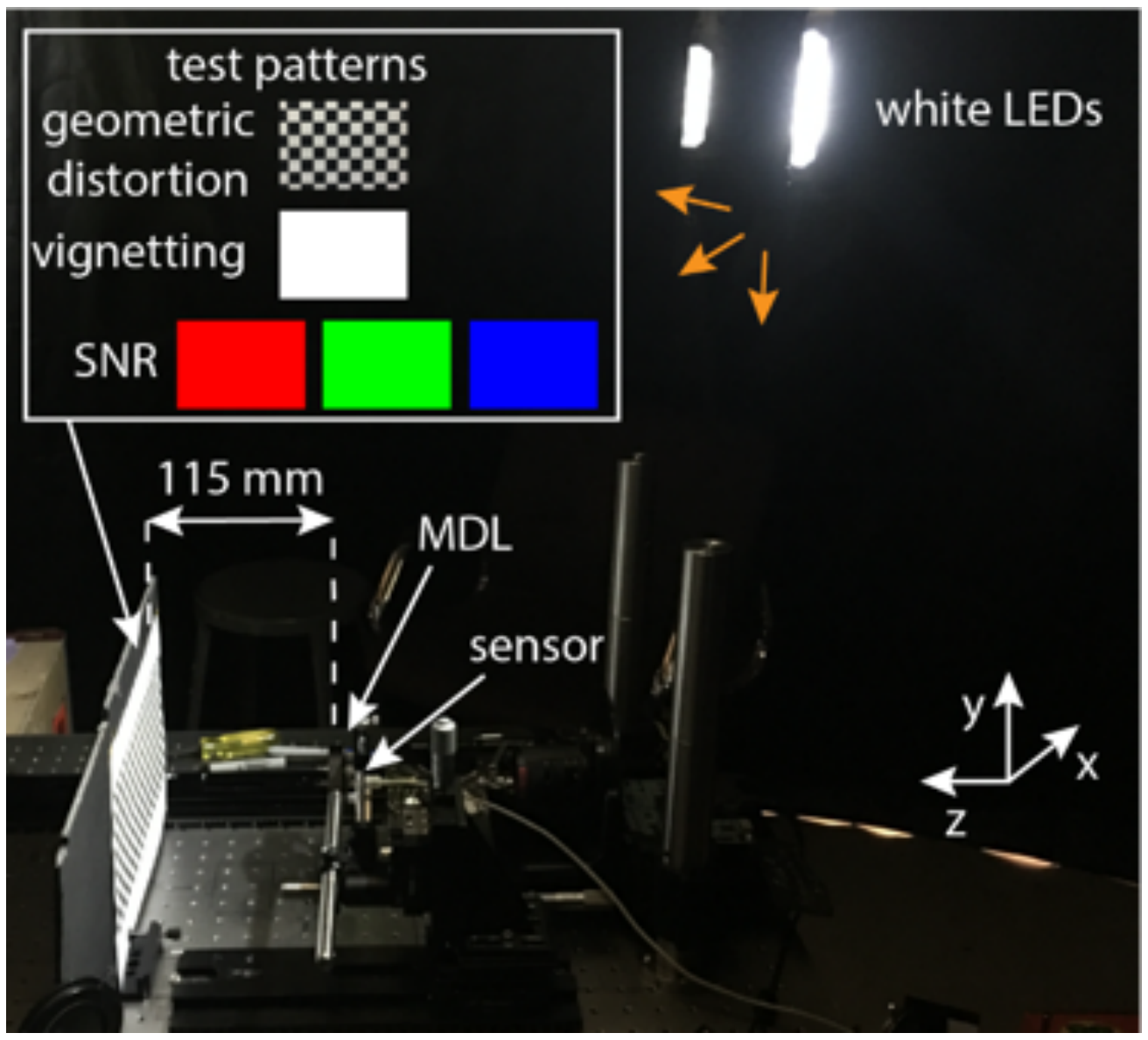

***Fig. S17:*** *Optical setup used to measure the geometric distortion, vignetting and signal-to-noise ratio (SNR) of a test VIS/NIR multi-level diffractive lens.*

**9. Illumination Spectrum Details**

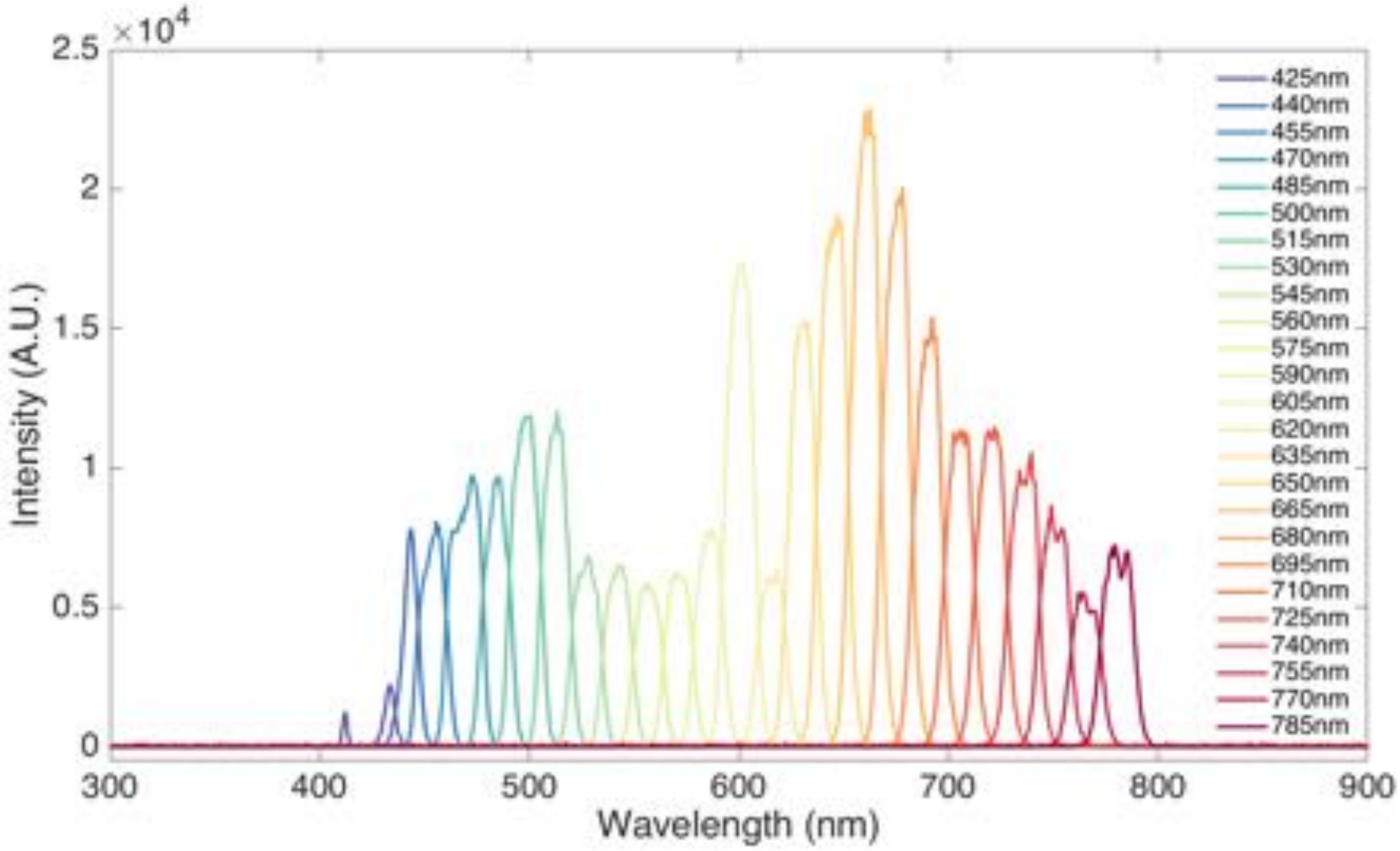

***Fig. S18:*** *Spectrum of SuperK Varia. For each wavelength, the bandwidth is 15 nm.The envelope of all these wavelengths is the full broadband.*

**10. MDL achromatic from 0.5mm to 15mm**

**10.1 Simulated scalar Point Spread Functions (PSFs)**

The scalar simulated point spread functions (PSFs) for the broadband lens designed in AZ9260 are provided below:

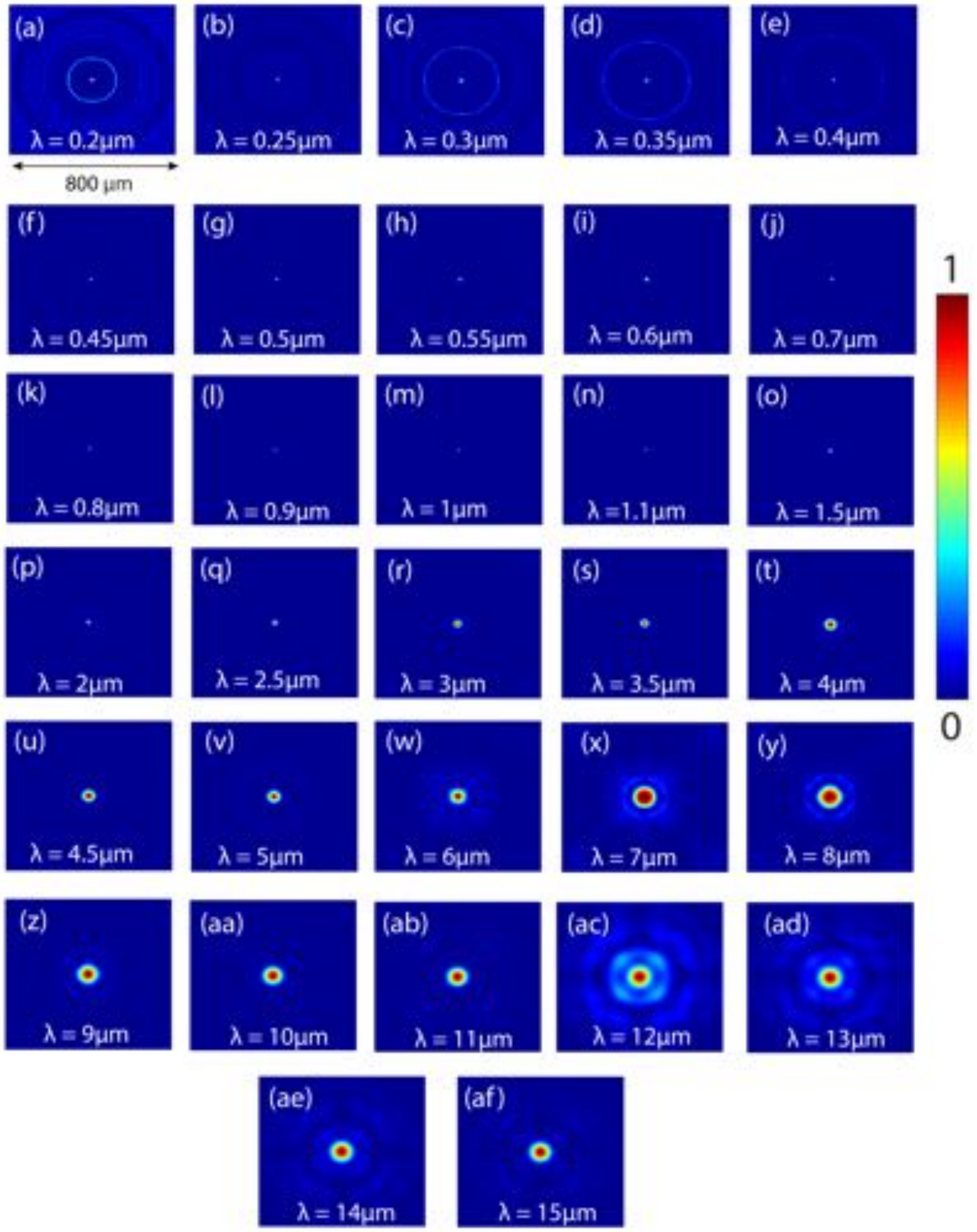

***Fig. S19:*** *Simulated PSFs for* ***(a)*** *0.2um* ***(b)*** *0.25um* ***(c)*** *0.3um* ***(d)*** *0.35um* ***(e)*** *0.4um* ***(f)*** *0.45um* ***(g)*** *0.5um* ***(h)*** *0.55um* ***(i)*** *0.6um* ***(j)*** *0.7um* ***(k)*** *0.8um* ***(l)*** *0.9um* ***(m)*** *1um* ***(n)*** *1.1um* ***(o)*** *1.5um* ***(p)*** *2um* ***(q)*** *2.5um* ***(r)*** *3um* ***(s)*** *3.5um* ***(t)*** *4um* ***(u)*** *4.5um* ***(v)*** *5um* ***(w)*** *6um* ***(x)*** *7um* ***(y)*** *8um* ***(z)*** *9um* ***(aa)*** *10um* ***(ab)*** *11um* ***(ac)*** *12um* ***(ad)****13um* ***(ae)****14um and* ***(af)****15um.*

**10.2. Simulated FWHM**

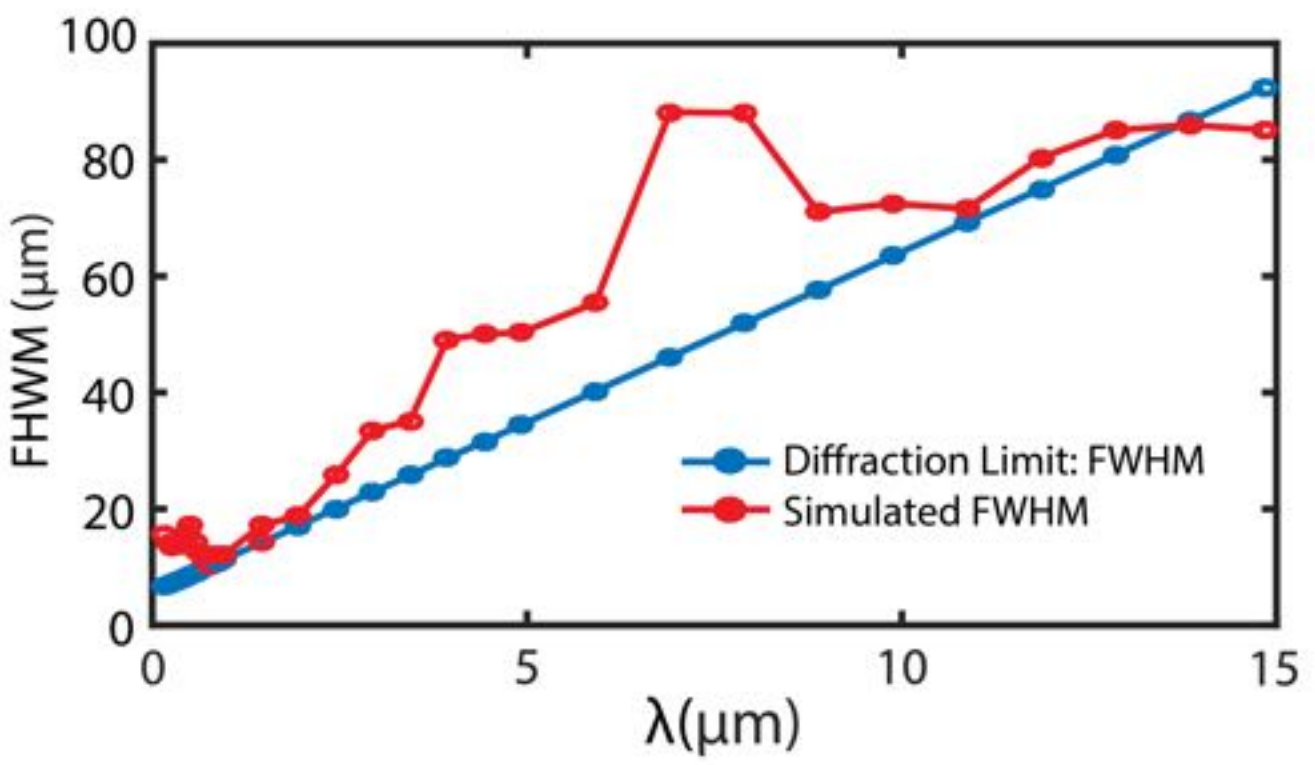

***Fig. S20:*** *FWHM. plot of the 0.5um – 15um broadband MDL.*

### 10.3. 2D FDTD Simulation Setup and Simulated z-propagation plot of the 0.5μm – 15μm broadband MDL

The 2D full wave FDTD simulations were carried out using Lumerical FDTD Solutions. The material properties (refractive index and absorption coefficient as a function of frequency) of the Photoresist (AZ9260) was imported into Lumerical directly as the structure's optical data. A ".lsf" script was written to replicate the lens geometry using the same dimensions, which was specified during the optimization process as depicted in Fig. S20(a). An incident plane wave (type: diffracting) along the backward "y-axis" direction illuminated the diffractive lens surface. For the broadband excitation, the entire range or bandwidth of the pulse was defined for the appropriate design.

The entire FDTD simulation region was considered from the back surface of the spherical lens right up to roughly ~1.5 times the distance from the focal plane. A Perfectly Matched Layer (PML) boundary condition set up. Due to the inherent symmetry of the designed structure, the x-min boundary was set to "anti-symmetric" which reduced the requirements by 1/2 of the original simulation requirements in terms of both time and memory. The default mesh was used to simulate the structures instead of a very fine mesh to avoid the huge computation time. The mesh accuracy was kept at "4" which has a good tradeoff for precision and accuracy versus the time and memory requirement. Field monitors placed to observe the field profiles of the propagating electromagnetic radiation.

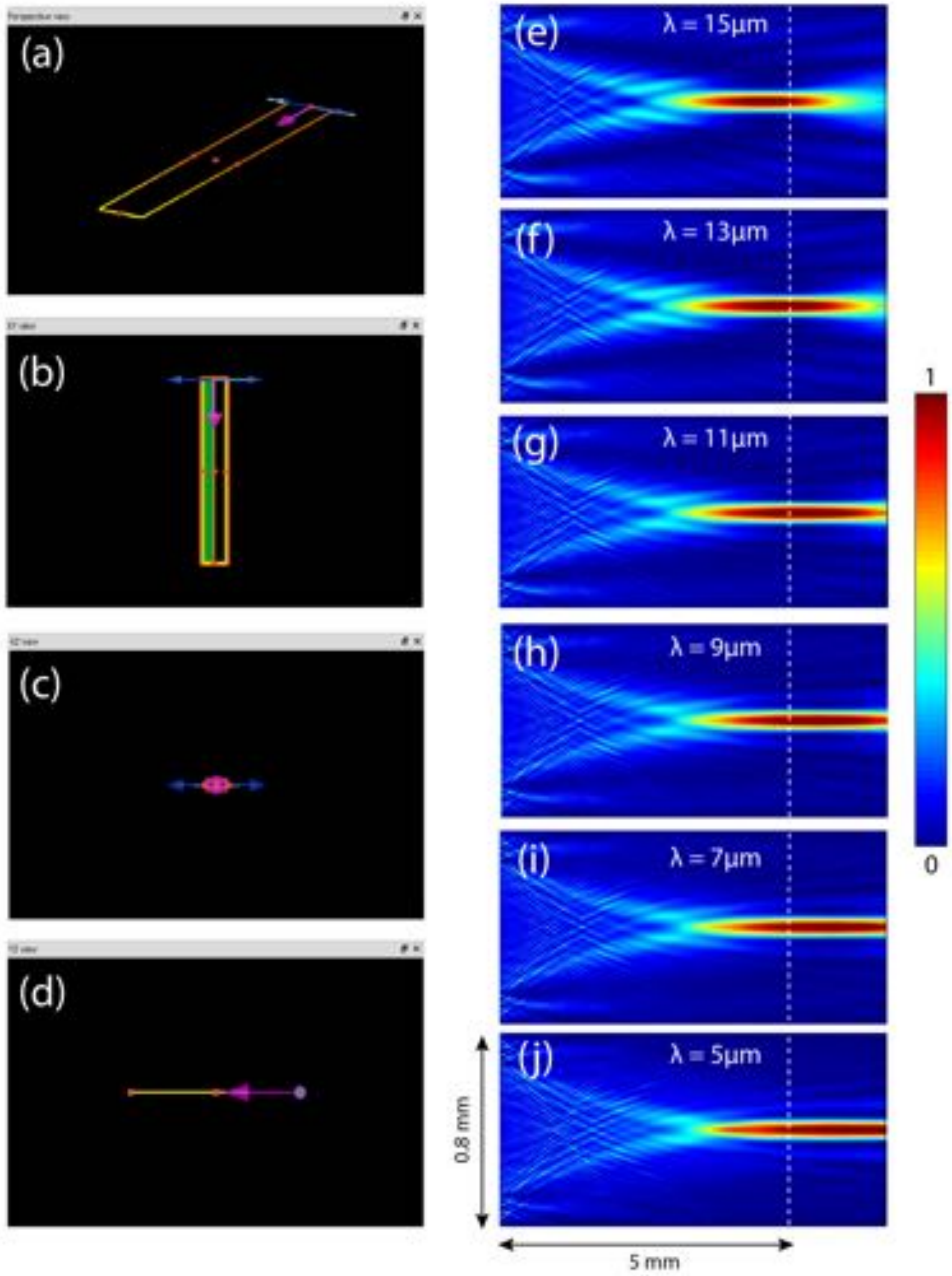

***Fig. S21:*** *2D FDTD simulation setup showing the (a) perspective view, (b) XY view, (c) XZ view and (d) YZ view in Lumerical FDTD solutions. Simulated z-propagation plot for* ***(e)*** *15um* ***(f)*** *13um* ***(g)*** *11um* ***(h)*** *9um* ***(i)*** *7um* ***(j)*** *5um.*

**11. Impact of degrees of freedom on 0.5μm to 15μm MDL performance.**

### 11.1. Effect of pixel width

We simulated the effect of both larger and smaller pixel width keeping the number of pixel height levels fixed at 100 in the design. We observed that the efficiency was almost constant across the range of 6um to 10 um; after which we see that the average efficiency fall by a margin of about ~15% when the pixel width was even increased to 12 um and finally at 14um the efficiency drops bellows 60%.

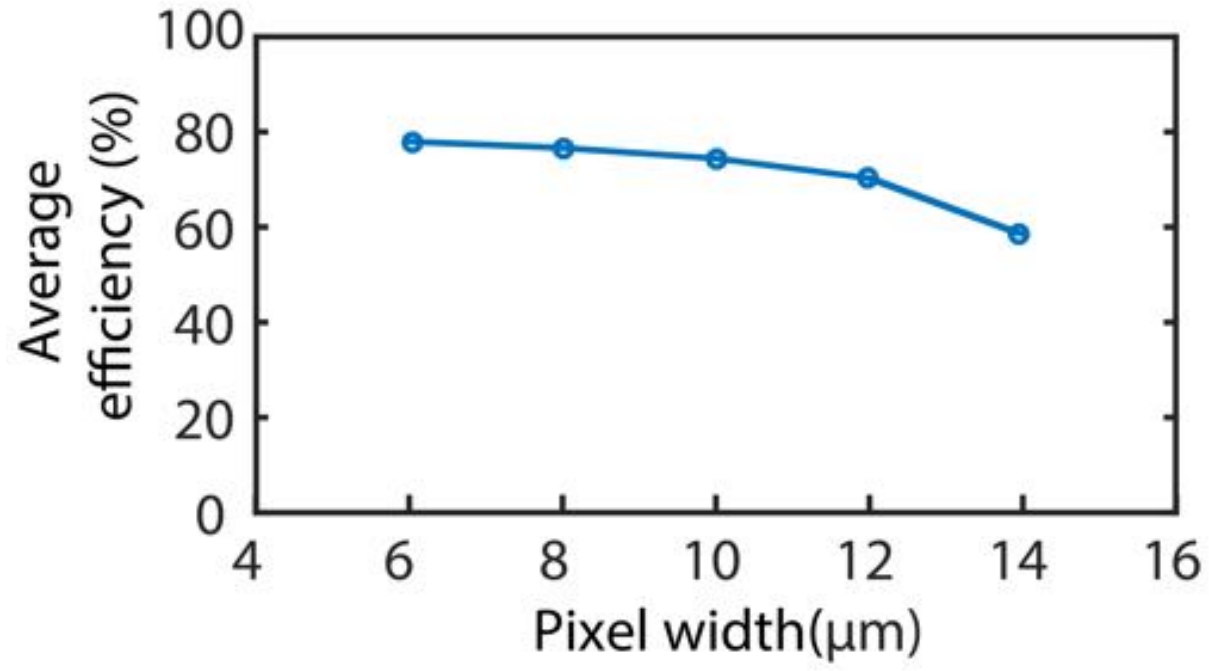

***Fig. S22:*** *Effect of pixel width on average efficiency of the 0.5um – 15um broadband MDL.*

### 11.2. Effect of number of height levels

We simulated the effect of both larger and smaller number of pixel height levels keeping the width of each individual pixel at 8um in the design. We observed that the efficiency increased by a substantial margin from 8 pixels levels to 64 pixels levels. After that, the increase the average efficiency was almost constant from 64 to 128 pixels. Hence, the use of 100 pixels was a good choice from a design perspective.

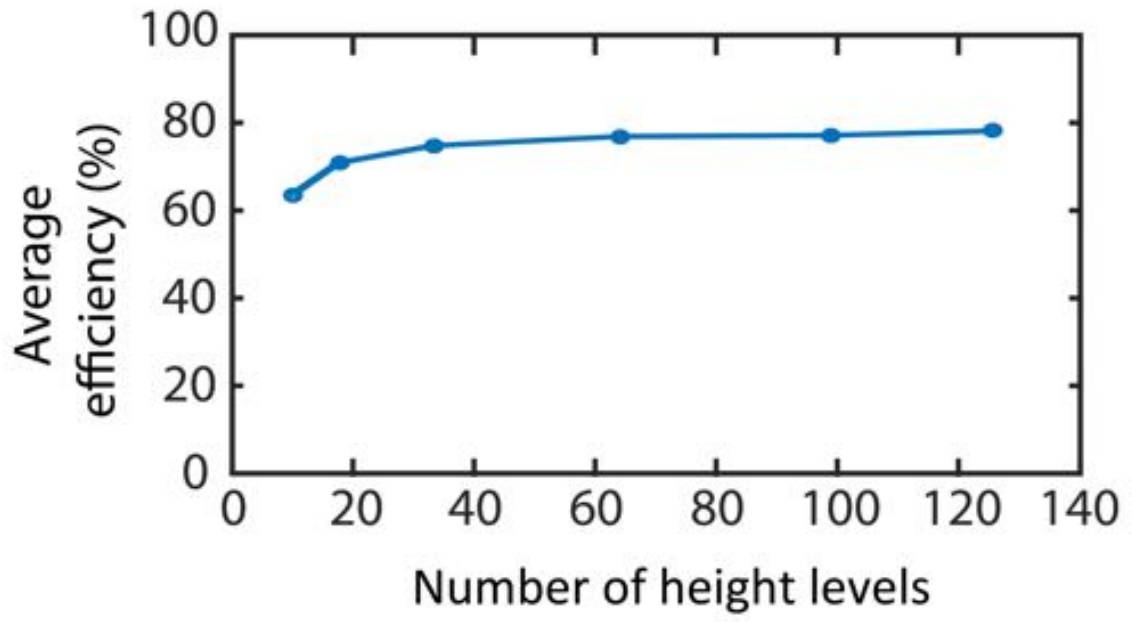

***Fig. S23:*** *Effect of the number of height levels on average efficiency of the 0.5um – 15um broadband MDL.*

## 12. Performance of an MDL achromatic over 2.5μm to 150μm.

### 12.1. Simulated scalar Point Spread Functions (PSFs)

The scalar simulated point spread functions (PSFs) for the broadband lens designed in silicon are provided below:

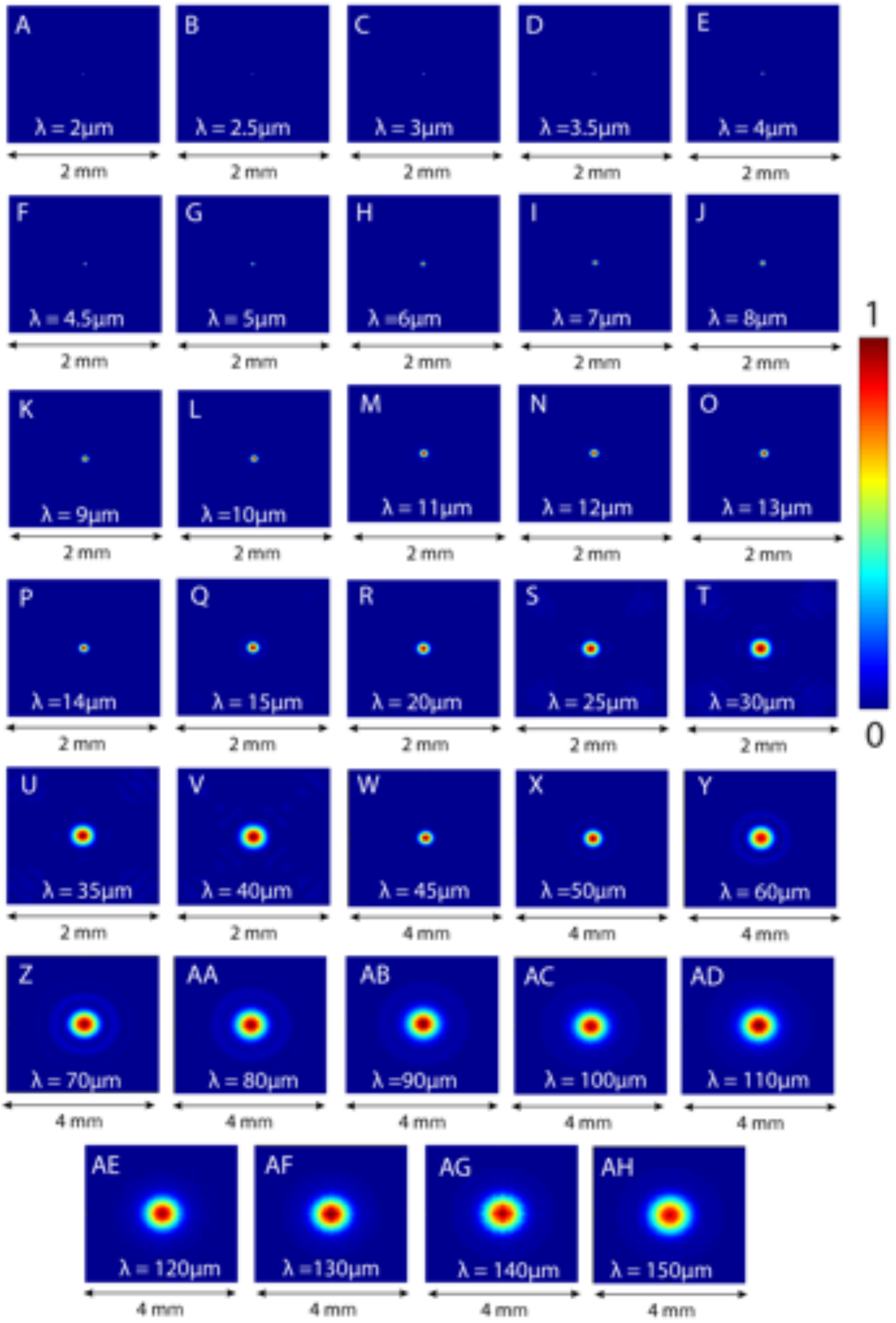

***Fig. S24:*** *Simulated PSFs for* ***(a)*** *2um* ***(b)*** *2.5um* ***(c)*** *3um* ***(d)*** *3.5um* ***(e)*** *4um* ***(f)*** *4.5um* ***(g)*** *5um* ***(h)*** *6um* ***(i)*** *7um* ***(j)*** *8um* ***(k)*** *9um* ***(l)*** *10um* ***(m)*** *11um* ***(n)*** *12um* ***(o)*** *13um* ***(p)*** *14um* ***(q)*** *15um* ***(r)*** *20um* ***(s)*** *25um* ***(t)*** *30um* ***(u)*** *35um* ***(v)*** *40um* ***(w)*** *45um* ***(x)*** *50um* ***(y)*** *60um* ***(z)*** *70um* ***(aa)*** *80um* ***(ab)*** *90um* ***(ac)*** *100um* ***(ad)****110um* ***(ae)****120um* ***(af)****130um* ***(ag)*** *140um and* ***(ah)*** *150um.*

**12.2. FWHM**

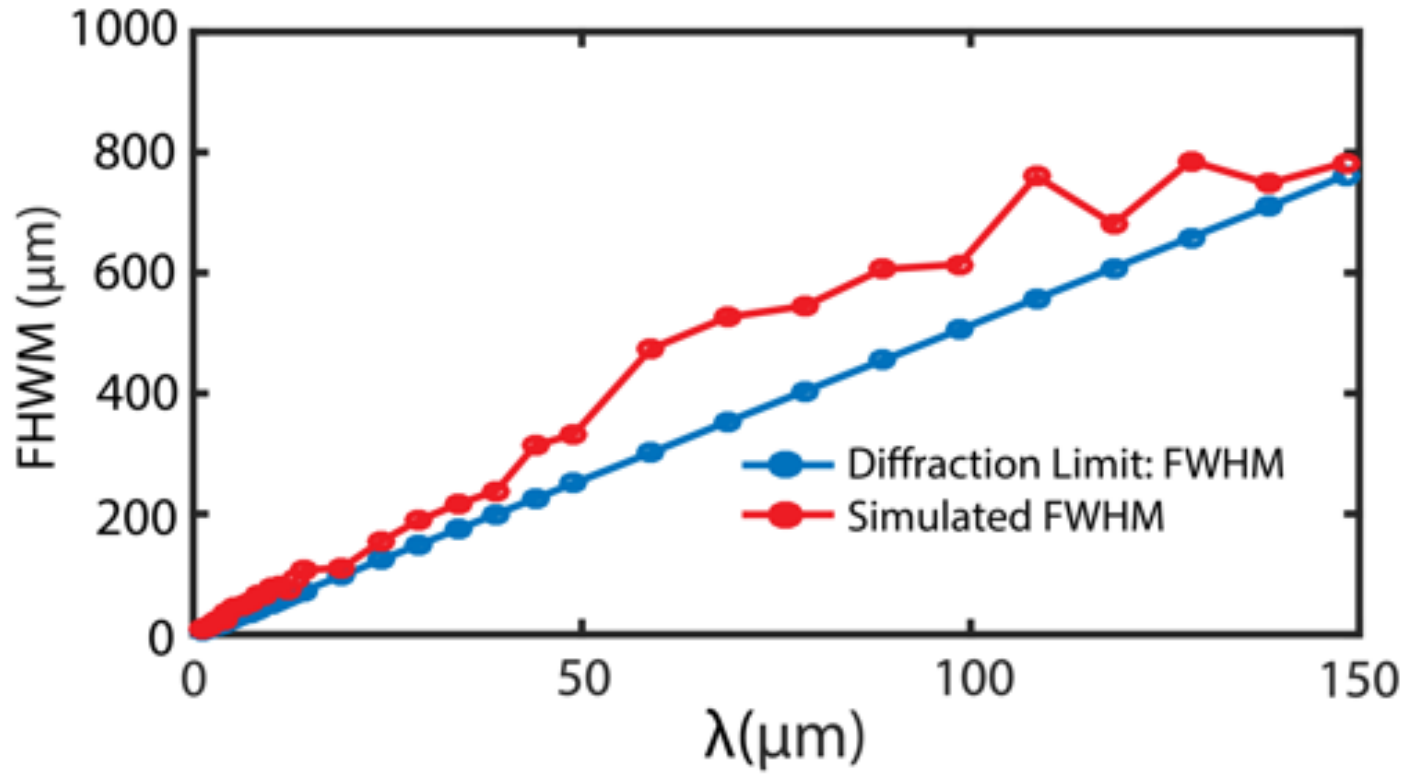

***Fig. S25:*** *FWHM plot of the 2.5um – 150um broadband MDL.*